\documentclass[%
pre,
reprint,
amsmath,amssymb,
aps,
]{revtex4-2}

\usepackage{graphicx}
\usepackage{dcolumn}
\usepackage{bm}
\usepackage[usenames]{xcolor}
\usepackage[normalem]{ulem}

\usepackage{array}
\usepackage{mathtools}
\usepackage{float}
\usepackage[boxruled]{algorithm2e}
\SetAlFnt{\footnotesize}
\SetAlCapFnt{\small}
\SetAlCapNameFnt{\small}
\usepackage{makecell}
\usepackage{booktabs}

\newcommand{\pin}{p_{\rm in}}
\newcommand{\pout}{p_{\rm out}}
\newcommand{\din}{d_{\rm in}}
\newcommand{\dout}{d_{\rm out}}

\newcommand{\calN}{\mathcal{N}}
\newcommand{\calE}{\mathcal{E}}
\newcommand{\calG}{\mathcal{G}}

\newcommand{\calS}{\mathcal{S}}
\newcommand{\calB}{\mathcal{B}}

\newcommand{\bfA}{\bm{A}}
\newcommand{\bfB}{\bm{B}}

\newcommand{\bfI}{\mathbf{I}}
\newcommand{\bfJ}{\mathbf{J}}
\newcommand{\bfD}{\bm{D}}
\newcommand{\bfC}{\mathbf{C}}
\newcommand{\bfM}{\mathbf{M}}
\newcommand{\bfN}{\bm{N}}

\newcommand{\bfQ}{\bm{Q}}
\newcommand{\bfU}{\mathbf{U}}
\newcommand{\bfV}{\mathbf{V}}

\newcommand{\bfBH}{\mathbf{BH}}
\newcommand{\bfOmega}{\bm{\Omega}}
\newcommand{\bfLambda}{\mathbf{\Lambda}}
\newcommand{\bfPsi}{\mathbf{\Psi}}
\newcommand{\bfTheta}{\mathbf{\Theta}}
\newcommand{\rmT}{\mathrm{T}}

\begin{document}
\title{Detectability of minority communities in networks}
\author{Jiaze Li}
\email{l.jiaze@maastrichtuniversity.nl}
\affiliation{Department of Data Analytics and Digitalisation, School of Business and Economics, Maastricht University}
\author{Leto Peel}
\email{l.peel@maastrichtuniversity.nl}
\affiliation{Department of Data Analytics and Digitalisation, School of Business and Economics, Maastricht University}
\date{\today}

\begin{abstract}
	Community structure is prevalent in real-world networks, with empirical studies revealing heterogeneous distributions where a few dominant majority communities coexist with many smaller groups. These small-scale groups, which we term \emph{minority communities}, are critical for understanding network organization but pose significant challenges for detection. Here, we investigate the detectability of minority communities from a theoretical perspective using the Stochastic Block Model. We identify three distinct phases of community detection: the \emph{detectable phase}, where overall community structure is recoverable but minority communities are merged into majority groups; the \emph{distinguishable phase}, where minority communities form a coherent group separate from the majority but remain unresolved internally; and the \emph{resolvable phase}, where each minority community is fully distinguishable. These phases correspond to phase transitions at the Kesten-Stigum threshold and two additional thresholds determined by the eigenvalue structure of the signal matrix, which we derive explicitly. Furthermore, we demonstrate that spectral clustering with the Bethe Hessian exhibits significantly weaker detection performance for minority communities compared to belief propagation, revealing a specific limitation of spectral methods in identifying fine-grained community structure despite their capability to detect macroscopic structures down to the theoretical limit.
\end{abstract}
\maketitle

\section{Introduction}
Community structure, characterized by groups of nodes that are more densely connected to each other than to the rest of the network, is prevalent in a wide range of real-world networks. 
These communities typically reveal important functional or organizational units within systems. 
These properties have drawn significant attention to community structure and its detection.
For instance, in brain functional networks, researchers have discovered modular structures that correspond to distinct communities~\cite{del2018finding}, such as those governing visual or auditory processing.
In social networks~\cite{fortunato2010community, nicolini2016modular}, connections among friends tend to be stronger and denser compared to those between strangers.
Understanding community structure provides a clearer view of network organization and enables a deeper analysis of the underlying systems.

Numerous methods have been developed to detect communities in networks~\cite{fortunato202220}, including algorithms based on modularity optimization~\cite{newman2004finding}, statistical inference~\cite{karrer2011stochastic}, and spectral clustering~\cite{saade2014spectral, dall2019revisiting, saade2016spectral}.
Essentially, these algorithms partition a network's nodes based on the differences in their connection patterns, e.g., identifying assortative communities with dense intra-community and sparse inter-community connections. 
Empirical studies of real-world networks identify heterogeneous distributions of community sizes, where a few dominant \emph{majority} communities coexist with many smaller groups~\cite{leskovec2009community}.
Examples of these include scientific collaboration networks~\cite{newman2004fast}, musician networks~\cite{gleiser2003community}, and email-based social networks~\cite{guimera2003self}.
In this study, we define these small-scale groups that comprise only a small fraction of the total node population as \emph{minority communities}.
Here, we explore the detection of these minority communities from a theoretical perspective, aiming to quantify the limits of their recovery.

Here we frame the problem of community detection as one of statistical inference. 
We assume that the observed network is generated from an underlying data generating process that depends on some latent community assignment.
A popular model for this process is the Stochastic Block Model (SBM)~\cite{holland1983stochastic}, in which edges are generated between nodes with a probability that is conditioned on their community memberships.
In this setting, the community detection problem can be viewed as recovering the latent community assignments from the relative density of connections among different node groups.
The detectability of communities relates to the noisy-channel coding theorem~\cite{shannon1948mathematical}, which states that information transmitted through a channel is unrecoverable if the channel's capacity is below the transmission rate.
For communities to be detectable, the signal provided by the community structure must be sufficiently strong relative to the noise introduced by random connections.
Communities are information-theoretically undetectable when the signal-to-noise ratio (SNR) of the relative edge densities falls below a certain threshold~\cite{abbe2016achieving}.
However, above this information-theoretic limit, there exists a \emph{computational} threshold known as the Kesten-Stigum bound~\cite{kesten1966additional}. 
In most cases, between these two thresholds lies a \emph{hard regime} where detection is information-theoretically possible but believed to be computationally intractable---no known polynomial-time algorithm can succeed~\cite{decelle2011asymptotic}.
Here, we consider this computational limit as the detectability threshold and anything below it as ``undetectable,'' meaning that no algorithm can recover the community structure with an accuracy better than random guessing.

The detectability threshold has been derived for a variety of different generative models, starting with the symmetric SBM---a model with equal-sized groups and uniform connection probabilities within and between communities~\cite{decelle2011asymptotic}. For this model, the Kesten-Stigum threshold marks the boundary below which efficient algorithms fail.
Abbe et al.~\cite{abbe2016achieving} derived an explicit expression for this threshold for a more general SBM, free from parameter restrictions. 
Comprehensive reviews of the research on detectability can be found in Abbe~\cite{abbe2018community} and Moore~\cite{moore2017computer}. 
Building on these foundational results, subsequent research has explored the detectability threshold for community structures in various other settings, 
including in networks that have hierarchical~\cite{peel2024detectability} or overlapping community structure~\cite{wu2021unraveling} and in directed~\cite{wilinski2019detectability} and dynamic networks~\cite{ghasemian2016detectability}.
Community detectability has also been considered in the context of hypergraphs~\cite{angelini2015spectral, chodrow2023nonbacktracking, ruggeri2024message, li2026higher}.
Further explorations into the detectability threshold of the SBM have uncovered intriguing phenomena.
For instance, Chin et al.~\cite{chin2025stochastic} find that when the number of communities is sufficiently large, detection may be possible even in the hard regime where $\mathrm{SNR}<1$.
Furthermore, the asymptotic analysis of Decelle et al.~\cite{decelle2011asymptotic} has been complemented by work from Young et al.~\cite{young2017finite}, who conducted a finite-size analysis of the detectability threshold.

Above the detectability threshold, efficient algorithms exist that can identify communities that correlate positively with the planted partition~\cite{massoulie2014community, mossel2014belief, abbe2016achieving}.
However, being above the detectability threshold does not guarantee that all communities are equally detectable.
A network may contain communities of varying sizes and connection densities, leading to heterogeneous recovery across different groups.
The weak recovery condition of detecting communities better than random guessing can be achieved through, for instance, the identification of a coarser partition that correlates with a finer planted partition~\cite{peel2024detectability}.
This distinction is critical: a network may possess a community structure that is considered detectable even if certain fine-grained groups remain indistinguishable. 
For instance, consider a network composed of two large communities and one small community. 
The small community may be too small or sparsely connected to be identified as a distinct community. 
However, an algorithm might still perform better than random guessing by overlooking the small group and absorbing its nodes into the two larger communities. 

This difficulty of detecting minority communities parallels the well-known resolution limit in modularity optimization~\cite{fortunato2007resolution}, which tends to overlook communities smaller than approximately $\sqrt{|\mathcal{E}|}$ and often merges them into larger groups.
Our observations show that even when $\mathrm{SNR}>1$, community detection methods, including those theoretically capable of reaching the detectability threshold, may still fail to identify smaller, \emph{minority} communities.
Nodes within these minority communities may either be uniformly distributed among majority communities or merged into a new aggregated community.
We derive explicit conditions that determine the detectability phase of such minority communities under spectral clustering. 
Furthermore, we demonstrate that spectral clustering exhibits significantly weaker detection performance for minority communities compared to belief propagation—an asymptotically optimal community detection method~\cite{ghasemian2016detectability}.
This indicates a specific limitation of spectral clustering in identifying minority communities, despite its ability to detect macroscopic community structures down to the theoretical detectability limit~\cite{saade2014spectral, krzakala2013spectral}.

In this work, we first review the detectability threshold for the stochastic block model. 
Building on this foundation, we introduce two specific SBM configurations that explicitly incorporate minority communities of relatively smaller size.
Subsequently, we present results on synthetically generated networks to demonstrate that some community detection methods systematically fail to identify minority communities, even when the network communities are above the detectability threshold.
We identify the existence of three distinct phases of detectability for minority communities and specify the conditions under which each phase occurs under spectral clustering using the Bethe Hessian.
Finally, we conclude with a discussion of our findings and suggest directions for future research.

\section{Detectability threshold for community detection}
Here we briefly introduce the stochastic block model (SBM) and the detectability threshold.
The SBM is a probabilistic generative model in which edges are generated conditionally independently, given the community assignments of the nodes.
For simplicity, we restrict our attention to undirected and unweighted networks, although the SBM can be easily extended to more general cases~\cite{Aicher_2014, wilinski2019detectability}.
We generate a network with $n$ nodes and $q$ communities using the SBM by first assigning each node $i$ to a community $\psi_i\in\{1, 2, ..., q\}$. 
Then, according to a $q\times q$ affinity matrix $\bfOmega$, for each pair of nodes $(i, j)$, we assign an edge between them with a probability $\bfOmega_{\psi_i \psi_j}$, which is determined by their respective community assignments.

The symmetric SBM, often referred to as the planted partition model~\cite{chaudhuri2012spectral, tsourakakis2015streaming, condon1999algorithms}, assumes all communities to be of equal size, such that each community contains the same number of nodes.
The affinity matrix $\bfOmega$ consists of two values: a diagonal value $\pin$, representing the intra-community edge probability, and an off-diagonal value $\pout$, representing the inter-community edge probability.
For a network generated by the symmetric SBM to be detectable, the difference between $\pin$ and $\pout$ must be sufficiently large. 
Formally, we can describe the detectability condition using the signal-to-noise ratio (SNR) written in terms of the expected number of edges within the same community, $\din=\frac{n}{q}\pin$, and to nodes in other communities, $\dout=\frac{n}{q}\pout$:
\begin{equation}\label{snr_symSBM_chap2}
    \mathrm{SNR}:=\frac{(\din-\dout)^2}{d} = 1\enspace ,
\end{equation}
where $d=\din+(q-1)\dout$ is the expected degree.

For a general SBM without any constraints on community sizes and the affinity matrix $\bfOmega$, its detectability threshold $\mathrm{SNR}=1$ is derived from the spectrum of the signal matrix $\bfQ$~\cite{abbe2016achieving, stephan2024community},
\begin{equation}
    \bfQ=\bfN\bfOmega \enspace,
\end{equation}
where $\bfN$ is a diagonal matrix with entries $N_{rr}$ representing the number of nodes in community $r$:
\begin{equation}
    \bfN_{rr}=|\left\{i \mid \psi_i=r\right\}|\enspace .
\end{equation}

Using the first- and second-largest eigenvalues of $\bfQ$, denoted as $\lambda_1$ and $\lambda_2$ respectively, the SNR for the general SBM is given by~\cite{abbe2018community, abbe2016achieving}:
\begin{equation}\label{snrgeneral_chap2}
    \mathrm{SNR}=\frac{\lambda_2^2}{\lambda_1}\enspace .
\end{equation}

The SNR expression in Eq.~\eqref{snrgeneral_chap2} for the general SBM reduces to that of the symmetric SBM.
In the symmetric SBM, we have $\bfN=\frac{n}{q}\bfI_q$ and $\bfOmega=(\pin-\pout)\bfI_q+\pout \bfJ_q$,
where $\bfI_q$ is the $q$-dimensional identity matrix and $\bfJ_q$ is the $q$-dimensional all-ones matrix.
Then the signal matrix becomes:
\begin{equation}
\bfQ=\frac{n}{q}((\pin-\pout)\bfI_q+\pout \bfJ_q) \enspace .
\end{equation}
Since the matrix $\bfQ$ takes the form $a\bfI_q+b\bfJ_q$, its eigenvalues ordered by non-increasing magnitude can be derived:
\begin{equation}
    \begin{aligned}
        &\lambda_1=\frac{n}{q}(\pin+(q-1)\pout)=\din+(q-1)\dout=d \\
        &\lambda_2=\frac{n}{q}(\pin-\pout) =\din-\dout
    \end{aligned} \enspace .
\end{equation}
Substituting these into equation~\eqref{snrgeneral_chap2} yields the SNR expression for the symmetric SBM as shown in~\eqref{snr_symSBM_chap2}.

\section{Detectability of minority communities} \label{chap3_minoritydetectability}
\begin{figure}
    \centering
    \includegraphics[width=\linewidth]{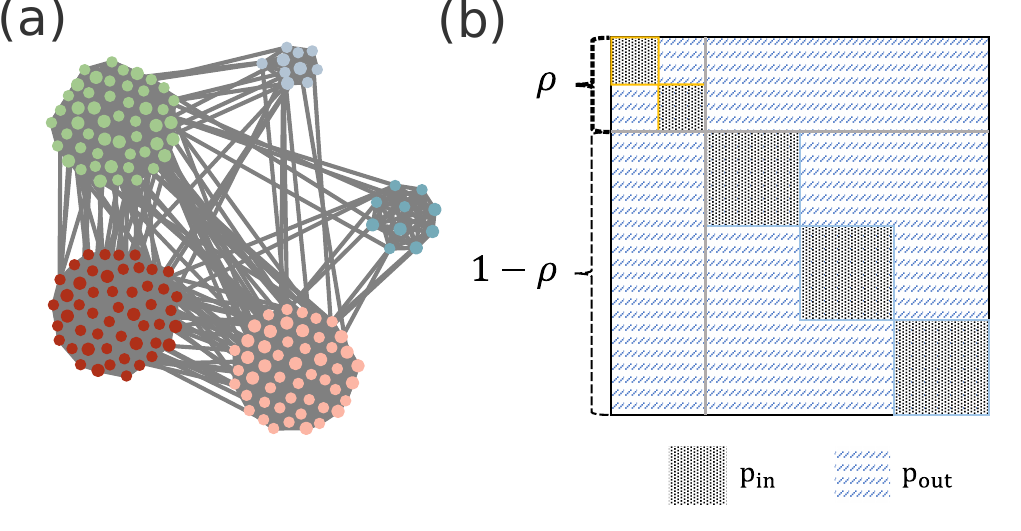}\\
	\caption{\small\it (a) A network with two minority communities and three majority communities.
    (b) An illustration of the expected adjacency matrix of the network shown in (a). Here, two minority communities share a total proportion of $\rho$ nodes equally, while three majority communities share the remaining $1-\rho$ proportion equally. The patterns in different blocks represent the expected values of adjacency matrix entries, corresponding to the edge probabilities $\pin$, $\pout^{(1)}$, and $\pout^{(2)}$ in the affinity matrix of the SBM. \label{fig3.1a}}
\end{figure}

In this section, we model minority communities using the SBM and derive the detectability thresholds for these communities under spectral clustering.
Figure~\ref{fig3.1a} illustrates the two community classes by size that we assume:
\begin{itemize}
    \item the minority communities $\cal S$ constitute a set of $q_s$ communities, each with fewer nodes than those in the majority class.
    \item the majority communities $\cal B$ constitute a set of $q_b$ communities with comparatively larger sizes,
\end{itemize}
where $q_s+q_b=q$.
Let $\rho$ denote the total proportion of nodes allocated to all minority communities in $\calS$, so each minority community holds $\frac{\rho}{q_s}$ proportion of nodes and each majority community holds $\frac{1-\rho}{q_b}$ proportion of nodes. If all $q$ communities were equally sized, each would hold $\frac{1}{q}$ of nodes, and the $q_s$ minority communities would collectively hold $\frac{q_s}{q}$. Since minority communities must be smaller than average, their total node share must fall below this equal-split baseline:
\begin{equation}\label{minority_condition}
\rho < \frac{q_s}{q} \enspace .
\end{equation}

We capture the per-community size discrepancy as $\epsilon$, the difference in node proportion between a minority and a majority community, 
\begin{equation}\label{epsilon<0}
    \epsilon=\frac{\rho}{q_s}-\frac{1-\rho}{q_b} \enspace ,
\end{equation}
which implies that $\epsilon<0$ to satisfy condition~\eqref{minority_condition} that minority communities are indeed smaller than majority communities.
Then the matrix of community sizes $\bfN$ is given by
\begin{equation}\label{minority_community_size_matrix}
    \begin{aligned}
        \bfN=n\cdot \mathrm{diag}\bigg(\bigg[\underbrace{\frac{\rho}{q_s}, \ldots, \frac{\rho}{q_s}}_{q_s}, \underbrace{\frac{1-\rho}{q_b}, \ldots, \frac{1-\rho}{q_b}}_{q_b}\bigg]\bigg)_{q\times q}\enspace
    \end{aligned} \enspace .
\end{equation}

Similar to the symmetric SBM, we specify the unique elements of the affinity matrix $\bfOmega$: $\pin$ for the intra-community edge probability, and $\pout$ for the inter-community edge probability. 
In this case, the affinity matrix $\bfOmega$ takes the form of the planted partition model:
\begin{equation}\label{MinorityConsistentPoutOmega}
    \bfOmega=(\pin-\pout)\bfI_q+\pout \bfJ_q\enspace ,
\end{equation}
and we can express the signal matrix $\bfQ = \bfN\bfOmega$ as a block matrix:
\begin{equation}\label{MinorityConsistentPoutNOmega}
\renewcommand{\arraystretch}{2}
    \bfQ
    = n \left[
    \begin{array}{c|c}
        \frac{\rho}{q_s}\bfOmega_{q_s}& \frac{\rho}{q_s}\pout \bfJ_{q_s\times q_b}  \\
        \hline
        \frac{1-\rho}{q_b}\pout \bfJ_{q_b\times q_s} & \frac{1-\rho}{q_b}\bfOmega_{q_b}\\
    \end{array}\right]_{q\times q}\enspace ,
\end{equation} 
where $\bfOmega_{q}$ is the planted partition affinity matrix from Eq.~\eqref{MinorityConsistentPoutOmega} with dimension $q$.

Each eigenvalue of $\bfQ$ exceeding $\sqrt{\lambda_1}$ in magnitude corresponds to an eigenvector encoding a detectable \emph{community contrast}---a bipartition of the
communities into two groups~\cite{krzakala2013spectral}. 
When $q$ such eigenvalues exist, collecting their eigenvectors into a $q$-dimensional embedding captures the full community structure, and spectral clustering in this space reliably distinguishes all communities down to the detectability threshold~\cite{bordenave2015non,abbe2015detection}.
When communities are of equal size ($\epsilon=0$), the signal matrix $\bfQ$ has only two distinct eigenvalues ($\lambda_1$ and a repeated $\lambda_2$ of multiplicity $q-1$), so the single condition $\mathrm{SNR}=\lambda_2^2/\lambda_1>1$ determines whether all communities are detectable.
With minority communities ($\epsilon < 0$), this degeneracy is broken: as we show below, the repeated eigenvalue splits into distinct values, each associated with a community contrast that distinguishes between minority and majority blocks, or within either block.

In the following subsections, we derive the eigenvalues of the signal matrix $\bfQ$ and label them with non-increasing magnitude: $\lambda_1\geq\lambda_2\geq\lambda_3\geq\lambda_4$. We then identify that the condition
\begin{equation}
    \frac{\lambda_r^2}{\lambda_1}=1, \quad r>2 \enspace ,
\end{equation}
which follows the form of the SNR in Eq.~\eqref{snrgeneral_chap2}, marks the phase transition boundaries for detecting each community contrast via spectral clustering with the Bethe Hessian matrix. 

\subsection{Eigenvalues of the signal matrix}
To identify the eigenvalues of the signal matrix, we work with the scaled matrix $\bfQ/n$, which is independent of network size and simplifies subsequent algebraic derivations. 
We make this change without loss of generality since the eigenvalue ratios determining detectability are invariant under scaling.
To find its eigenvalues, we find the roots of the characteristic polynomial $\det(\bfQ/n - \lambda \bfI_q) = 0$. 
This determinant can be expressed using the block structure:
\begin{equation}\label{chap2_blockmatrix}
    \renewcommand{\arraystretch}{1.8}
    \begin{aligned}
         \det & \left(\frac{\bfQ}{n}-\lambda \bfI_q\right)\\
        &=  \det\left(\left[
        \begin{array}{cc}
            \frac{\rho}{q_s} \bfOmega_{q_s}-\lambda \bfI_{q_s} & \frac{\rho}{q_s} \pout \bfJ_{q_s\times q_b}\\
            \frac{1-\rho}{q_b}\pout \bfJ_{q_b\times q_s} & \frac{1-\rho}{q_b}\bfOmega_{q_b}-\lambda \bfI_{q_b}
        \end{array}
        \right]\right)\\
        &= \det\left(\left[
        \begin{array}{cc}
            \hat{\bfA} & \hat\bfB\\
            \hat\bfC & \hat\bfD
        \end{array}
        \right]\right) \enspace .
    \end{aligned}
\end{equation}

According to the determinant formula for a block matrix~\cite{silvester2000determinants}, if $\hat\bfA$ is invertible, then
\begin{equation}\label{blockmatrixdeterminate}
    \det\left(\left[
        \begin{array}{cc}
            \hat\bfA & \hat\bfB\\
            \hat\bfC & \hat\bfD
        \end{array}
        \right]\right)=\det(\hat\bfA)\det(\hat\bfD-\hat\bfC\hat\bfA^{-1}\hat\bfB)\enspace .
\end{equation}

We derive the eigenvalues of the scaled signal matrix $\bfQ/n$, labeling them in 
non-increasing order as $\lambda_1 \geq \lambda_2 \geq \lambda_3 \geq \lambda_4$. 
Our derivation proceeds by exploiting the block structure: we identify the contributing eigenvector from the minority submatrix $\frac{\rho}{q_s} \bfOmega_{q_s}$, and then we identify the remaining eigenvalues by solving for $\lambda \neq \lambda_4$.

The submatrix $\frac{\rho}{q_s} \bfOmega_{q_s}$ has two distinct eigenvalues: one 
corresponding to the \emph{uniform} eigenvector (all entries equal within the minority block), and one corresponding to community contrast vectors. Only the latter extends to a full-matrix eigenvalue with multiplicity $q_s-1$:
\begin{equation}\label{lambda_4_definition}
    \lambda_4=\frac{\rho}{q_s}\left(\pin-\pout\right)=\frac{\rho}{q_s}\delta \enspace ,
\end{equation}
such that $\delta = \pin - \pout$ denotes the difference between intra- and inter-community connection probabilities. 
The eigenvalue corresponding to the uniform eigenvector only extends to the full matrix under specific conditions (see Appendix~\ref{appendix_5_lambda1_hat_bfA}).

For $\lambda \neq \lambda_4$, we solve
\begin{equation}\label{characteristiceq}
    \det(\hat\bfD-\hat\bfC\hat\bfA^{-1}\hat\bfB)=0 \enspace ,
\end{equation}
as specified in equation~\eqref{blockmatrixdeterminate} to determine the remaining eigenvalues of $\bfQ/n$. Expanding Eq.~\eqref{characteristiceq} (full details in Appendix~\ref{appendix_6}), we obtain:
\begin{equation}\label{detD-CA-1B}
    \det(\hat\bfD-\hat\bfC\hat\bfA^{-1}\hat\bfB) =
    \left(\frac{1-\rho}{q_b}\delta-\lambda\right)^{\!q_b-1}
    \frac{Q(\lambda)}{\frac{\rho}{q_s}\delta+\rho\pout-\lambda} \enspace ,
\end{equation}
where
\begin{align}\label{Qdef}
    Q(\lambda) = &
    \left(\frac{1-\rho}{q_b}\delta-\lambda\right)\!
    \left(\frac{\rho}{q_s}\delta+\rho\pout-\lambda\right) \notag \\
    & + (1-\rho) \pout\left(\frac{\rho}{q_s}\delta-\lambda\right) .
\end{align}

From Eq.~\eqref{detD-CA-1B}, we identify one root as
\begin{equation}
    \lambda_2=\frac{(1-\rho)}{q_b}\delta \enspace ,
\end{equation}
with multiplicity $q_b-1$. This eigenvalue corresponds to contrasts between majority communities. The remaining two roots, $\lambda_3$ and $\lambda_1$, are obtained by solving for $Q(\lambda) = 0$ in equation~\eqref{Qdef}. Rewriting this factor as a standard quadratic equation
$a\lambda^2+b\lambda+c=0$, the coefficients are
\begin{align}\label{coefficients}
    a&=1 \enspace , \notag \\
    b 
    &=-\left(\frac{\rho}{q_s}+\frac{1-\rho}{q_b}\right)\delta-\pout \enspace , \notag \\
    c 
    &=\frac{\rho(1-\rho)\delta^2}{q_bq_s}+\rho(1-\rho)\pout\delta\left(\frac{1}{q_s}+\frac{1}{q_b}\right) \enspace .
\end{align}

The two roots correspond to $\lambda_1$ and $\lambda_3$:
\begin{equation}
    \lambda_1, \lambda_3 = 
        \frac{1}{2}\left[\left(\frac{\rho}{q_s}+\frac{1-\rho}{q_b}\right)\delta+\pout \pm \sqrt{\Delta}\,\right] \enspace ,
\end{equation}
where 
\begin{equation}
    \Delta =\pout^2+\epsilon^2\delta^2+2(2\rho-1)\pout\delta\epsilon \enspace,
\end{equation}
and $\epsilon=\frac{\rho}{q_s}-\frac{1-\rho}{q_b}$ (as previously defined in Eq.~\eqref{epsilon<0}). 
The larger of the two, $\lambda_1$, is the largest eigenvalue that corresponds with the \emph{global uniform} eigenvector (the all-ones vector across all communities). 
The roots coincide, i.e., $\lambda_1=\lambda_3$, when $\Delta=0$. 
Using the inequality $x^2 + y^2 \geq 2|xy|$:
\begin{align}
    \Delta &\geq  2|\pout\epsilon\delta|+2(2\rho-1)\pout\epsilon\delta \label{AM_GM_eq}\\
    &= \left\{
    \begin{array}{cc}
        4\rho\pout\epsilon\delta >0    &,\text{if $\delta<0$}   \\
        0 &, \text{if $\delta=0$} \\
        4(\rho-1)\epsilon\delta >0 &, \text{if $\delta>0$}
    \end{array}\right. \enspace ,
\end{align}
with the inequality in Eq.~\eqref{AM_GM_eq} saturated when $|\pout|=|\epsilon\delta|$. Then, $\lambda_1=\lambda_3$ occurs only if $\pin=\pout=0$.

In the assortative case, these eigenvalues satisfy $\lambda_1 \geq \lambda_2 \geq \lambda_3 \geq \lambda_4 \geq 0$ and comprise:
\begin{itemize}
    \item Eigenvalue $\lambda_1 = \frac{1}{2}\left[\left(\frac{\rho}{q_s}+\frac{1-\rho}{q_b}\right)\delta+\pout+ \sqrt{\Delta}\right]$ with multiplicity 1, corresponding to the \emph{global uniform} eigenvector (the all-ones vector across all communities).
    \item Eigenvalue $\lambda_2 = \frac{1-\rho}{q_b}\delta$ with multiplicity $q_b-1$, representing contrasts between majority communities.
    \item Eigenvalue $\lambda_3 = \frac{1}{2}\left[\left(\frac{\rho}{q_s}+\frac{1-\rho}{q_b}\right)\delta+\pout- \sqrt{\Delta}\right]$ with multiplicity 1, representing mixed contrasts between minority and majority groups.
    \item Eigenvalue $\lambda_4 = \frac{\rho}{q_s}\delta$ with multiplicity $q_s-1$, representing contrasts between minority communities.
\end{itemize}

The total count $1 + (q_b-1) + 1 + (q_s-1) = q_b + q_s$  matches the dimension of $\bfQ/n$. These eigenvalues are further verified using an alternative derivation in Appendix~\ref{appendix_7_anotherderiving}.
Each distinct eigenvalue corresponds to a specific community contrast vector. The detectability of each contrast is determined by whether its respective SNR exceeds unity, i.e., $\lambda_r^2/\lambda_1 > 1$.

In the assortative case ($\delta>0$), we verify that these eigenvalues satisfy $\lambda_1 \geq \lambda_2 \geq \lambda_3 \geq \lambda_4 \geq 0$. First, we analyze the sign of each eigenvalue. Note that:
\begin{align}
    &\left(\frac{\rho}{q_s}+\frac{1-\rho}{q_b}\right)\delta+\pout \notag\\
    =&\left(\frac{\rho}{q_s}+\frac{1-\rho}{q_b}\right)\pin+\left(1-\frac{\rho}{q_s}-\frac{1-\rho}{q_b}\right)\pout \notag \\
    \geq& 0 \enspace ,
\end{align}
which implies $\lambda_1\geq0$. The signs of $\lambda_4$ and $\lambda_2$ are also positive in the assortative case.
To determine the sign of $\lambda_3$, consider the product:
\begin{equation}
    \begin{aligned}
        \lambda_1\lambda_3=&\frac{1}{4}\left[ \left(\left(\frac{\rho}{q_s}+\frac{1-\rho}{q_b}\right)\delta+\pout\right)^2-\right.\\
        &\left.(\epsilon\delta+\pout)^2 -  4(\rho-1)\epsilon\delta\pout \,\right]\\
        =&\frac{1}{4}\left[ 4\frac{\rho(1-\rho)}{q_sq_b}\delta^2+4\frac{1-\rho}{q_b}\pout\delta-4(\rho-1)\epsilon\pout\delta \right]\\
        =&\frac{\rho(1-\rho)}{q_sq_b}\delta^2+\rho(1-\rho)\left(\frac{1}{q_s}+\frac{1}{q_b}\right)\pout\delta\\
        =&\frac{\rho(1-\rho)}{q_sq_b}\left[\delta+(q_s+q_b)\pout\right]\delta
    \end{aligned}\enspace .
\end{equation}
Given that $\delta>0$, the product $\lambda_1\lambda_3>0$. Since $\lambda_1\geq0$, it follows that $\lambda_3\geq 0$.

We now compare the eigenvalues:
\begin{align*}
    \lambda_1-\lambda_2
    &= \frac{1}{2}\left[\epsilon\delta+\pout + \sqrt{\Delta}\,\right]\\
    &\geq \frac{1}{2}\Bigl(\epsilon\delta+\pout + |\epsilon\delta+\pout|\Bigr)\geq0 \\
    \lambda_3-\lambda_2
    &= \frac{1}{2}\left[ \epsilon\delta+\pout-\sqrt{\Delta}\,\right]\\
    &\leq \frac{1}{2}\Bigl[\epsilon\delta+\pout - |\epsilon\delta+\pout|\Bigr]\leq0 \\
    \lambda_3-\lambda_4
    &= \frac{1}{2}\left[ -\epsilon\delta+\pout-\sqrt{\Delta}\,\right]\\
    &\geq \frac{1}{2}\Bigl[ - \epsilon\delta+\pout - |\epsilon\delta-\pout|\Bigr]\geq0
 \enspace ,
\end{align*}
which confirms that the eigenvalues are ordered as:
\begin{equation}
    \label{eq_eigval_order}
    \lambda_1\geq \lambda_2 \geq \lambda_3 \geq \lambda_4 \enspace .
\end{equation}

When $q_s=1$, the eigenvalue $\lambda_4$ (multiplicity $q_s-1$) vanishes. Similarly, when $q_b=1$, the eigenvalue $\lambda_2$ (multiplicity $q_b-1$) vanishes.
The eigenvalues of signal matrix $\bfQ$ are simply $n$ times these values. 
In these cases we relabel the eigenvalues in terms of their ordinal value. 
We summarize the eigenvalue formulas for different cases of $q_s$ and $q_b$ in Table~\ref{PQlambda}.
Blank entries indicate vanishing eigenvalues (multiplicity zero).
\begin{table}[H]
    \centering
    \renewcommand{\arraystretch}{2}
    \begin{tabular}{ccccc}
         \makecell{$\epsilon<0$\\ and $0<\rho<\frac{q_s}{q_s+q_b}$} & \makecell{$q_s>1$\\$q_b>1$} & \makecell{$q_s=1$\\$q_b>1$} & \makecell{$q_s>1$\\$q_b=1$} & \makecell{$q_s=1$\\$q_b=1$} \\
         \toprule
         $\frac{n}{2}\left[\left(\frac{\rho}{q_s}+\frac{1-\rho}{q_b}\right)\delta+\pout+ \sqrt{\Delta}\,\right]$ & $\lambda_1$ & $\lambda_1$& $\lambda_1$ & $\lambda_1$ \\
         \midrule
         $n\frac{1-\rho}{q_b}\delta$ & $\lambda_2$ & $\lambda_2$ &  & \\
         \midrule
         $\frac{n}{2}\left[\left(\frac{\rho}{q_s}+\frac{1-\rho}{q_b}\right)\delta+\pout- \sqrt{\Delta}\,\right]$ & $\lambda_3$ & $\lambda_3$ & $\lambda_2$ & $\lambda_2$\\
         \midrule
         $n\frac{\rho}{q_s}\delta$ & $\lambda_4$ & & $\lambda_3$ & \\
         \bottomrule
    \end{tabular}
    \caption{Eigenvalue formulas of signal matrix $\bfQ$ for different cases of $q_s\geq1$ and $q_b\geq1$. Here $\Delta=\pout^2+\epsilon^2\delta^2+2(2\rho-1)\pout\delta\epsilon$ and $\epsilon=\frac{\rho}{q_s}-\frac{1-\rho}{q_b}$. Blank entries indicate vanishing eigenvalues (multiplicity zero).}
    \label{PQlambda}
\end{table}

\subsection{Phase transition thresholds for detectability}\label{conjectured_pt}
The eigenvalue structure reveals two critical phase transitions governing community detectability. These thresholds follow directly from the SNR condition $\mathrm{SNR}=\lambda_r^2/\lambda_1>1$ established in Eq.~\eqref{snrgeneral_chap2}, where each eigenvalue $\lambda_r$ corresponds to a distinct community contrast.

The first transition occurs when:
\begin{equation}\label{minority_distinguishable_thres}
    \frac{\lambda_3^2}{\lambda_1}=1 \enspace ,
\end{equation}
marking the onset of the \textbf{distinguishable phase}, beyond which minority communities become spectrally distinguishable from the majority as a collective entity. At this threshold, the signals from all minority communities merge into one coherent mode, collectively separated from the majority but not yet separated from each other. This condition also emerges naturally when merging all minority communities into a single group and comparing against the merged majority, where the SNR for that configuration coincides with $\lambda_3^2/\lambda_1$.

The second transition occurs when:
\begin{equation}\label{minority_resolvable_thres}
    \frac{\lambda_4^2}{\lambda_1} =1 \enspace ,
\end{equation}
marking the onset of the \textbf{resolvable phase}, beyond which each minority community becomes spectrally resolvable as a distinct entity. At this threshold, each minority community develops its own independent signal, allowing clustering algorithms to resolve them separately from both the majority and from one another.

\begin{figure}
    \centering
    \includegraphics[width=\linewidth]{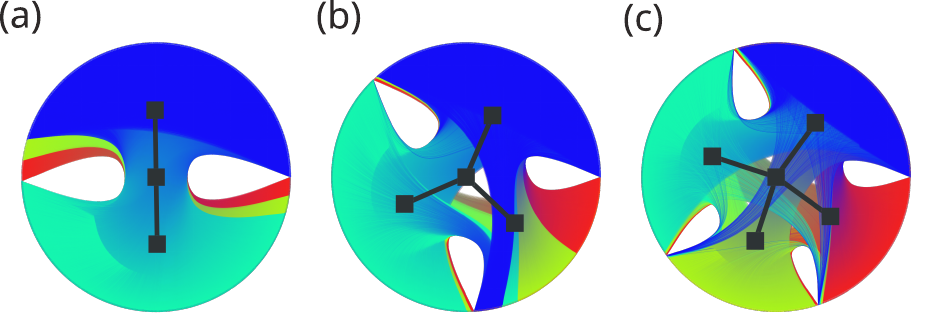}\\
	\caption{\small\it Schematic illustration of the three phases for a network with $q_s=2$ minority and $q_b=2$ majority communities: panel (a) shows the \textbf{detectable} phase where minorities are indistinguishable from majorities ($q=2$); panel (b) shows the phase where minorities are \textbf{distinguishable} but not yet fully resolved ($q=3$); panel (c) shows the phase where all communities are \textbf{resolvable} ($q=4$). \label{fig3.3}}
\end{figure}

Figure~\ref{fig3.3} schematically illustrates the three phases for a network with $q_s=2$ minority and $q_b=2$ majority communities: panel (a) shows the \textbf{detectable phase} ($q=2$), where minorities are indistinguishable from majorities; panel (b) shows the \textbf{distinguishable phase} ($q=3$), where minorities form a coherent group separate from majorities but remain unresolved internally; panel (c) shows the \textbf{resolvable phase} ($q=4$), where all communities are resolved separately.

These phase transition boundaries are verified empirically in Section~\ref{sec_spectral_experiments}, where spectral clustering with the Bethe Hessian matrix exhibits sharp transitions at precisely these theoretical thresholds, confirming both the distinguishable and resolvable phases.

\section{Detecting minority communities using spectral clustering}\label{sec_spectral_experiments}
To verify the theoretical phase transitions derived above, we use spectral clustering based on the Bethe Hessian matrix, which is known to achieve optimal detection performance down to the theoretical detectability threshold~\cite{saade2014spectral}.
The Bethe Hessian matrix can be derived from the non-backtracking matrix~\cite{saade2016spectral}, offering a linearized formulation of belief propagation~\cite{krzakala2013spectral}.
For a network with adjacency matrix $\bfA$ and degree matrix $\bfD$, which is a diagonal matrix with node degrees as the diagonal elements, the Bethe Hessian matrix $\bfBH_\eta$ is defined as:
\begin{equation}\label{chap3_bethehessian}
    \bfBH_\eta = (\eta^2-1)\mathbf{I}+ \mathbf{D}-\eta \mathbf{A} \enspace ,
\end{equation}
where $\eta$ is the regularization parameter, typically chosen such that $|\eta|=\sqrt{d}$, with $d$ being the average degree of the network~\cite{saade2014spectral}. This choice helps separate informative eigenvalues as negative values.
The spectral clustering procedure using the Bethe Hessian matrix is detailed in Algorithm~\ref{scbh}.
\begin{algorithm}
    \caption{Bethe Hessian Spectral Clustering.}
    \label{scbh}
    \SetAlgoLined
    \KwData{Adjacency matrix $\bfA$.}
    \KwResult{Community assignments matrix $\bfPsi$.}    
    Compute diagonal degree matrix $\bfD$\;
    Compute average degree $d$\;
    Compute $\eta=\sqrt{d}$\;
    Compute $\bfBH_{\pm \eta}$ according to~\eqref{chap3_bethehessian}\;
    Compute spectral decompositions $\bfBH_{\eta}=\bfV\bfLambda \bfV^\rmT$ and $\bfBH_{-\eta}=\bfU\bfTheta \bfU^\rmT$, where $\bfLambda=\mathrm{diag}(\lambda_1, \lambda_2, ...)$, $\bfTheta=\mathrm{diag}(\theta_1, \theta_2, ...)$\;
    Compute $q_+\leftarrow|\left\{\lambda_i:\lambda_i\leq 0\right\}|$ and $q_-\leftarrow|\left\{\theta_i:\theta_i\leq 0\right\}|$\;
    Estimate number of communities $q=q_++q_-$\;
    Select eigenvectors $\bfV^{q_+}$ and $\bfU^{q_-}$ with non-positive eigenvalues\;
    Form matrix $\left[\bfV^{q_+}, \bfU^{q_-}\right]$ by stacking these eigenvectors column-wise\;
    Run k-means clustering on the rows of $\left[\bfV^{q_+}, \bfU^{q_-}\right]$:                 \\$\bfPsi\leftarrow \textrm{k-means}(\left[\bfV^{q_+}, \bfU^{q_-}\right], q)$\;
    \Return $\bfPsi$;
\end{algorithm}

Notably, the order selection method for Bethe Hessian spectral clustering involves counting the number of negative eigenvalues of $\bfBH_{\sqrt{d}}$, which is simple.
To accommodate both assortative and disassortative community structures, we consider the number of negative eigenvalues for both $\bfBH_{\sqrt{d}}$ and $\bfBH_{-\sqrt{d}}$. 
Here we only consider assortative communities, but we provide additional results in Appendix~\ref{appendix_9} to show that these phases naturally extend to disassortative communities.

Next, we compare the two thresholds defined in equations~\eqref{minority_distinguishable_thres} and~\eqref{minority_resolvable_thres} with experimental results obtained using Bethe-Hessian-based spectral clustering.
We generate networks for each parameter pair $(\rho, \delta)$ and present the average $\rm AMI$ value along with the number of detected communities over multiple independent repetitions.
The network generation process follows the procedure detailed in Appendix~\ref{appendix_8_network_generation}.
Here, we use the Adjusted Mutual Information~\cite{vinh2009information} ($\rm AMI$) measure to evaluate the quality of detected communities.
For detected community assignments $\bm{\psi^*}$ and planted community assignments $\bm{\psi}$, the mutual information $\mathrm{MI}(\bm{\psi}, \bm{\psi^*})$ measures their statistical dependence:
\begin{equation}
    \begin{aligned}
        &\mathrm{MI}(\bm{\psi}, \bm{\psi^*})\\
        &=\sum_{r=1}^q\sum_{s=1}^{q^*}P(\psi_i=r, \psi^*_i=s)\log\frac{P(\psi_i=r, \psi^*_i=s)}{P(\psi_i=r)P(\psi^*_i=s)}
    \end{aligned} \enspace .
\end{equation}

The $\rm AMI$ corrects for chance using a random permutation null model that preserves the marginal community sizes:
\begin{equation}\label{ami}
    \begin{aligned}
        &\mathrm{AMI}(\bm{\psi}, \bm{\psi^*}) \\
        =& \frac{ \mathrm{MI}(\bm{\psi}, \bm{\psi^*}) - \mathbb{E}[\mathrm{MI}(\bm{\psi}, \bm{\psi^*})] }{ \mathrm{max}(\mathcal{H}(\bm{\psi}), \mathcal{H}(\bm{\psi^*})) - \mathbb{E}[\mathrm{MI}(\bm{\psi}, \bm{\psi^*})] }
    \end{aligned} \enspace ,
\end{equation}
where $\mathcal{H}(\bm{\psi})=-\sum_{r=1}^{q}P(\psi_i=r)\log P(\psi_i=r)$ is the entropy of $\bm{\psi}$, quantifying its uncertainty.
The $\rm AMI$ score ranges from 0 to 1, where 0 indicates that the detected communities are no better than a random assignment, and 1 corresponds to a perfect match with the true community structure.

\begin{figure}
    \centering
    \includegraphics[width=\linewidth]{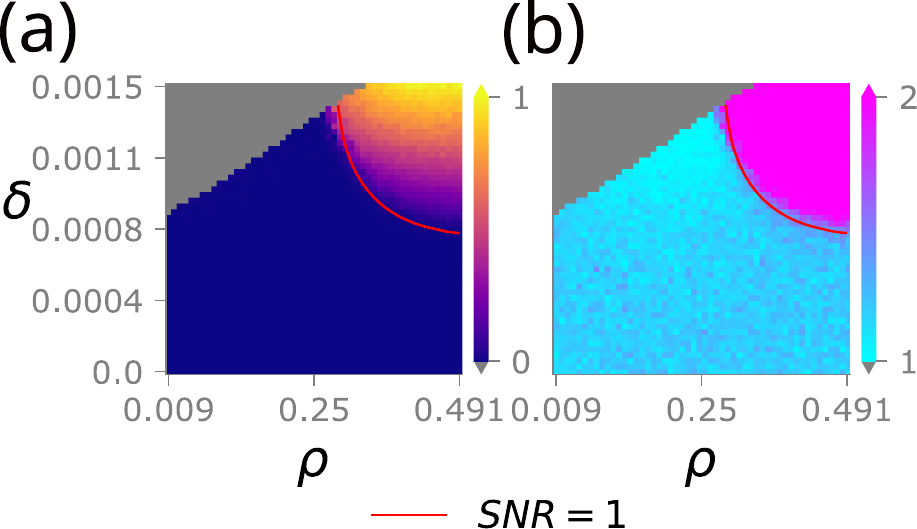}\\
	\caption{\small\it Experimental results under the consistent $\pout$ scenario with $n=6000$, $d=5$, $q_s=1$, $q_b=1$.
    For each $(\rho, \delta)$, the experiment was repeated 50 times. Community detection was performed using Bethe-Hessian-based spectral clustering on networks generated under the corresponding parameter pair.
    The gray area indicates invalid $(\rho, \delta)$ values that result in $\pin, \pout\notin[0, 1]$.
    (a) The average $\rm AMI$ of detected communities.
    (b) The average number of detected communities. Both (a) and (b) are shown alongside the curve $\mathrm{SNR}=1$. \label{fig3.5}}
\end{figure}

First, we analyze the simplest case, $q_s=1$, $q_b=1$.
According to Table~\ref{PQlambda}, the signal matrix has only two distinct eigenvalues in this case.
Figure~\ref{fig3.5} shows that spectral clustering successfully detects two communities only when $\rm SNR>1$;
below this threshold, all nodes are grouped into a single community.
No additional phase transitions are observed beyond $\rm SNR=1$.

\begin{figure}
    \centering
    \includegraphics[width=\linewidth]{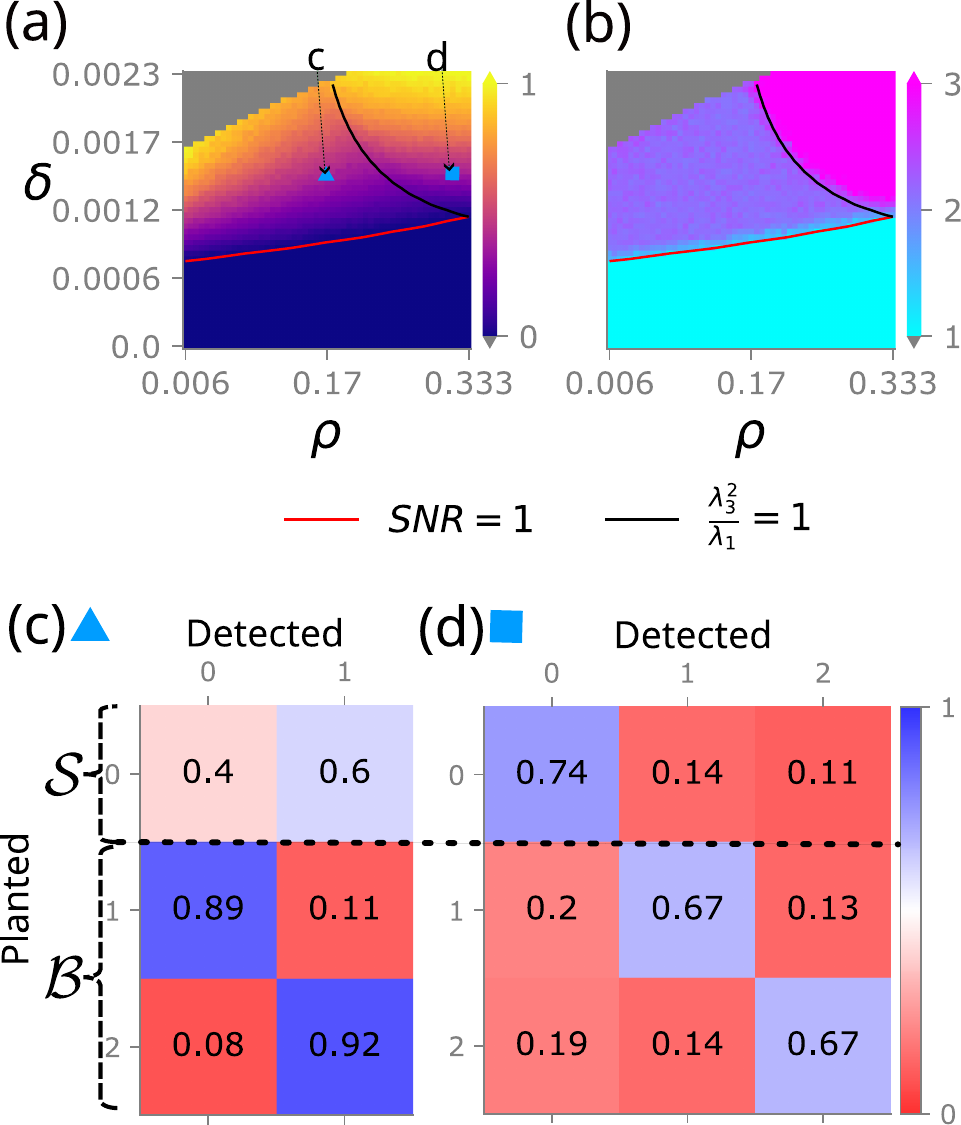}\\
	\caption{\small\it Experimental results under the consistent $\pout$ scenario with $n=6000$, $d=5$, $q_s=1$, $q_b=2$. For each $(\rho, \delta)$, the experiment was repeated 50 times.
    (a) The average $\rm AMI$. (b) The average number of detected communities. Both (a) and (b) are shown with curves $\mathrm{SNR}=1$ and $\frac{\lambda_3^2}{\lambda_1}=1$. (c, d) The confusion matrices at two $(\rho, \delta)$ points: (c) $(0.17, 0.0014)$, (d) $(0.32, 0.0014)$. \label{fig3.6}. In these matrices, each row corresponds to a true community, each column to a detected community, and the value at position $(r, s)$ indicates the proportion of nodes from true community $r$ assigned to detected community $s$.}
\end{figure}

Second, we consider the case with more than one majority community by setting $q_s=1, q_b=2$. Above $\mathrm{SNR}=1$, a second phase transition is predicted to occur at the curve $\frac{\lambda_3^2}{\lambda_1}=1$, where we expect the number of detected communities to increase from two to three. To characterize how the community structure evolves across this transition, we examine two representative $(\rho, \delta)$ points on either side of the curve. For each point, we compute the distribution of nodes from each planted community across the detected communities. These distributions are visualized as confusion matrices in Figure~\ref{fig3.6}(c, d), where rows correspond to planted communities, columns to detected communities, and entries $(r, s)$ represent the proportion of nodes from planted community $r$ assigned to detected community $s$.

Panel (b) in Figure~\ref{fig3.6} shows that the number of detected communities increases from two to three as we predicted as $(\rho, \delta)$ crosses the black curve defined by $\frac{\lambda_3^2}{\lambda_1}=1$. The confusion matrices in panels (c) and (d) clarify how the community structure differs between these two regimes. In Figure~\ref{fig3.6}(c), where $\mathrm{SNR}>1$ but $\frac{\lambda_3^2}{\lambda_1}<1$, the nodes of the minority community (labeled 0) are distributed relatively evenly between the two detected majority communities (labeled 1 and 2), indicating that the minority remains indistinguishable from the majority in this phase. By contrast, in Figure~\ref{fig3.6}(d), where $\frac{\lambda_3^2}{\lambda_1}>1$, all three communities are successfully detected, showing that exceeding this threshold allows us to clearly distinguish the minority community as a distinct entity from the majority.

\begin{figure}
    \centering
    \includegraphics[width=\linewidth]{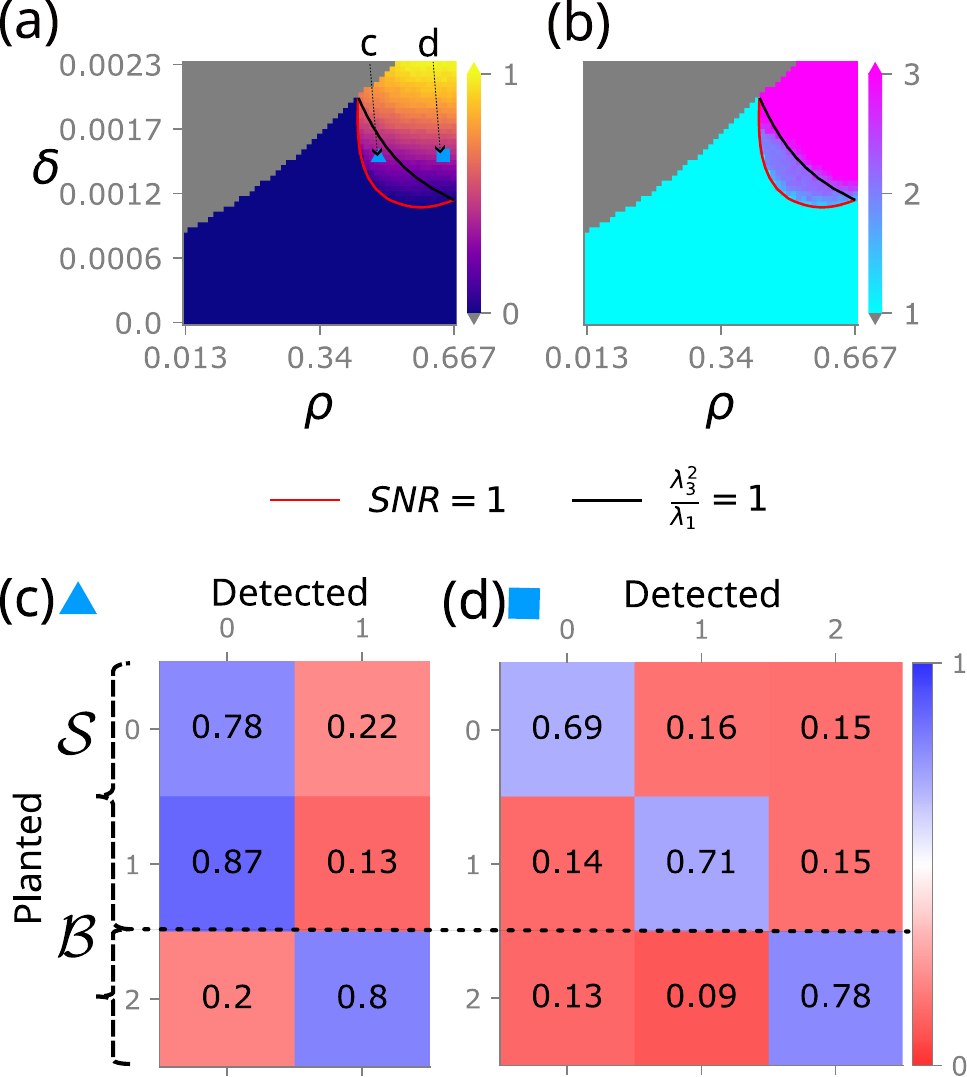}\\
	\caption{\small\it Experimental results under the consistent $\pout$ with $n=6000$, $d=5$, $q_s=2$, $q_b=1$. For each $(\rho, \delta)$, the experiment was repeated 50 times. (a) The average $\rm AMI$ of detected communities. (b) The average number of detected communities. Both (a) and (b) are shown with curves $\mathrm{SNR}=1$ and $\frac{\lambda_3^2}{\lambda_1}=1$. (c, d) The confusion matrices at two $(\rho, \delta)$ points: (c) $(0.5, 0.0014)$, (d) $(0.65, 0.0014)$. \label{fig3.7}}
\end{figure}

Third, we examine the case where we have more than one minority community with a single majority community, setting $q_s=2, q_b=1$.
The results, shown in Figure~\ref{fig3.7}, reveal an additional phase transition occurring above $\rm SNR=1$, which is also closely approximated by the curve $\frac{\lambda_3^2}{\lambda_1}=1$.
However, unlike the previous scenario, the case with multiple minority communities produces qualitatively different behavior when $\mathrm{SNR}>1$ but $\frac{\lambda_3^2}{\lambda_1}<1$.
The confusion matrices in Figure~\ref{fig3.7}(c, d) show the difference. In Figure~\ref{fig3.7}(c), where $\mathrm{SNR}>1$ but $\frac{\lambda_3^2}{\lambda_1}<1$, the two minority communities are merged into a single detected community---the minority communities are distinguishable from the majority yet irresolvable from each other. In contrast, Figure~\ref{fig3.7}(d), where $\frac{\lambda_3^2}{\lambda_1}>1$, shows all three communities correctly separated, confirming that the minority communities are fully resolvable above this threshold.

\begin{figure}
    \centering
    \includegraphics[width=\linewidth]{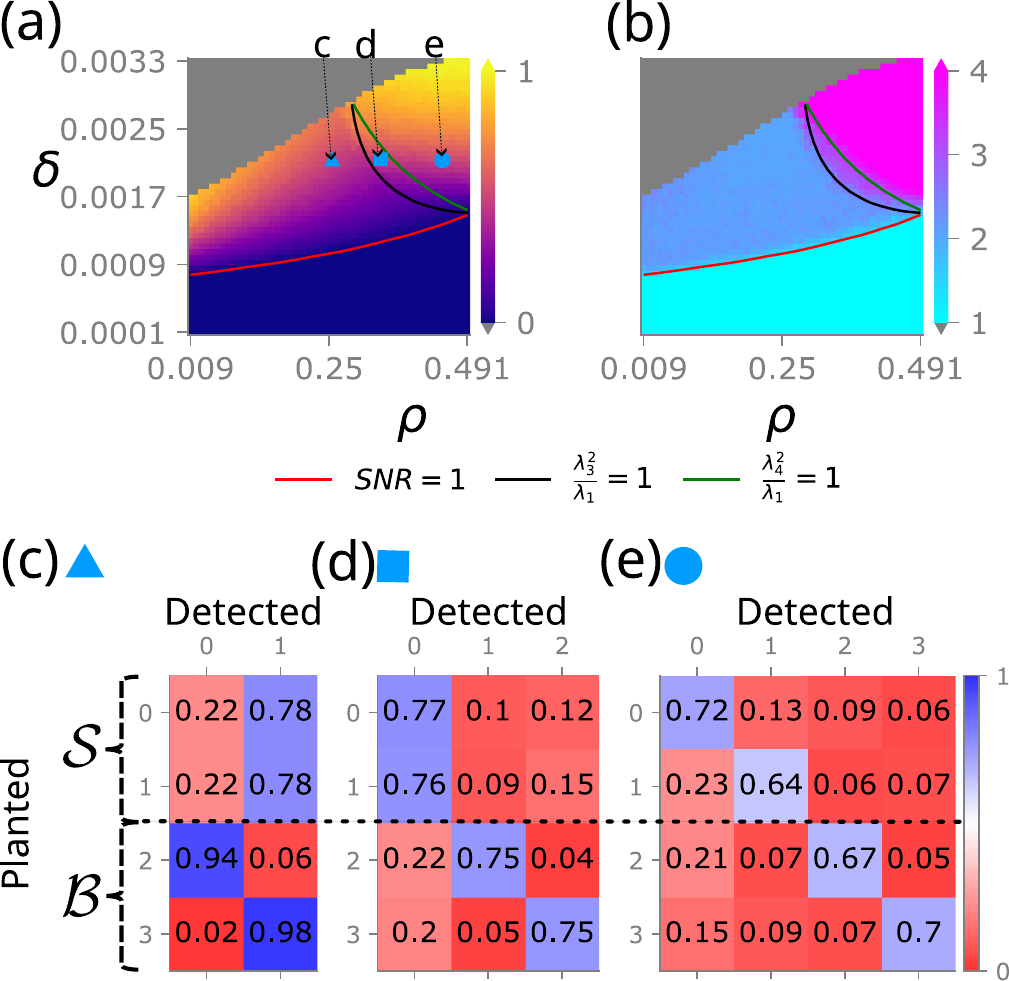}\\
	\caption{\small\it Experimental results with $n=6000$, $d=5$, $q_s=2$, $q_b=2$. For each valid $(\rho, \delta)$, the experiment was repeated 50 times. (a) The AMI of detected communities. (b) The number of detected communities. Both (a) and (b) are shown with curves $\mathrm{SNR}=1$, $\frac{\lambda_3^2}{\lambda_1}=1$, and $\frac{\lambda_4^2}{\lambda_1}=1$. (c--e) The confusion matrices at three $(\rho, \delta)$ points: (c) $(0.25, 0.0019)$, (d) $(0.39, 0.0019)$, (e) $(0.44, 0.0019)$. \label{fig3.8}}
\end{figure}

Finally, we examine a case with multiple minority and majority communities coexisting $(q_s=q_b=2)$.
Table~\ref{PQlambda} predicts four distinct eigenvalues, corresponding to two phase transitions above $\mathrm{SNR}=1$: the first at $\frac{\lambda_3^2}{\lambda_1}=1$ and the second at $\frac{\lambda_4^2}{\lambda_1}=1$.
Figure~\ref{fig3.8} confirms these predictions: the number of detected communities increases from two to three across the black curve, and from three to four across the green curve.
The confusion matrices in Figure~\ref{fig3.8}(c--e) show how minority detectability evolves across these phases. In Figure~\ref{fig3.8}(c), where $\mathrm{SNR}>1$ but $\frac{\lambda_3^2}{\lambda_1}<1$, the minority communities remain indistinguishable from the majority, with their nodes distributed evenly between the two detected majority communities. Furthermore, Figure~\ref{fig3.8}(d), where $\frac{\lambda_3^2}{\lambda_1}>1$ but $\frac{\lambda_4^2}{\lambda_1}<1$, shows all minority communities merged into a single detected community---distinguishable from the majority yet irresolvable from each other. Figure~\ref{fig3.8}(e), where $\frac{\lambda_4^2}{\lambda_1}>1$, displays all four communities correctly separated, confirming full resolvability above the second threshold.

\section{Comparison of detection methods}\label{sec_comparison_detection_methods}

In this section, we compare five combinations of community detection and
order-selection methods for detecting minority communities in networks
generated by the stochastic block model. We focus on two detection algorithms:
Bethe Hessian spectral clustering (BH)~\cite{saade2014spectral}, using Algorithm~\ref{scbh}, and belief propagation (BP)~\cite{decelle2011asymptotic}. BP is known to achieve asymptotically
optimal accuracy down to the detectability threshold~\cite{zdeborova2016statistical}. 
We pair these two detection methods with three order-selection strategies: negative-eigenvalue counting (NEC), minimizing free-energy (MFE), and minimum description length (MDL).

Order selection is important here because, unlike in standard detectability analysis where $q$ treated as known, minority community detection involves identifying different phases corresponding to detecting distinct numbers of communities: all communities separately, the minorities combined separate from the majority  communities, or the minorities merged into majority groups. 
The choice of order determines which phases are observed. 
We present results under two experimental settings: consistent inter-community edge probability and consistent average degree.

\subsection{Order-selection methods}\label{sec_order_selection}

We select the order using the NEC method by simply counting the number of negative eigenvalues of the Bethe Hessian matrix $\bfBH_{\sqrt{d}}$ (and $\bfBH_{-\sqrt{d}}$ for disassortative structures) to estimate $q$. This approach is simple and requires no additional computation beyond spectral decomposition.

The MFE method involves $q$ by minimizing free energy of the SBM using the Bethe approximation~\cite{decelle2011asymptotic}. We iteratively learn the SBM parameters $\bfN$ and $\bfOmega$ using expectation-maximization (EM)~\cite{dempster1977maximum} for each candidate $q$, compute the minimum free energy $f_q^{\rm min}$, and select the smallest $q$ that minimizes $f_q^{\rm min}$:
\begin{equation}
    q^*= \mathrm{argmin}_{q}(f_q^{\rm min}) \enspace .
\end{equation}
The relationship between $q$ and $f_q^{\rm min}$ typically exhibits an elbow shape. Determining $q^*$ requires comparing successive $f_q^{\rm min}$ values (full details in Appendix~\ref{appendix_4_round_free_energy_for_elbow_point}).

The MDL method allows us to select $q$ by minimizing the description length~\cite{peixoto2013parsimonious}, which quantifies the amount of information required to encode both the network data and the model parameters:
\begin{equation}\label{mdl_findq*}
    q^*= \mathrm{argmin}_{q}(\mathrm{L}) \enspace .
\end{equation}

For the SBM, the description length L is defined as
\begin{equation}
    \mathrm{L} = |\calE| h\left(\frac{q(q+1)}{2|\calE|}\right)+n\log(q)-|\calE|\sum_{rs}m_{rs}\log\left(\frac{m_{rs}}{n_rn_s}\right) \enspace ,
\end{equation}
where $h(x)=(1+x)\log(1+x)-x\log(x)$, $|\calE|$ is the number of edges in the network, $m_{rs}=\frac{|\calE_{rs}|}{|\calE|}$ is the proportion of edges between community $r$ and $s$, and $n_r=\frac{|\calN_r|}{n}$ denotes the proportion of nodes in community $r$.

\subsection{Consistent inter-community edge probability}
Using the same experimental setup as Section~\ref{chap3_minoritydetectability}, we apply combinations of community detection and order selection methods to networks generated with varying values of $\rho$ and $\delta$.
\begin{figure}
    \centering
    \includegraphics[width=\linewidth]{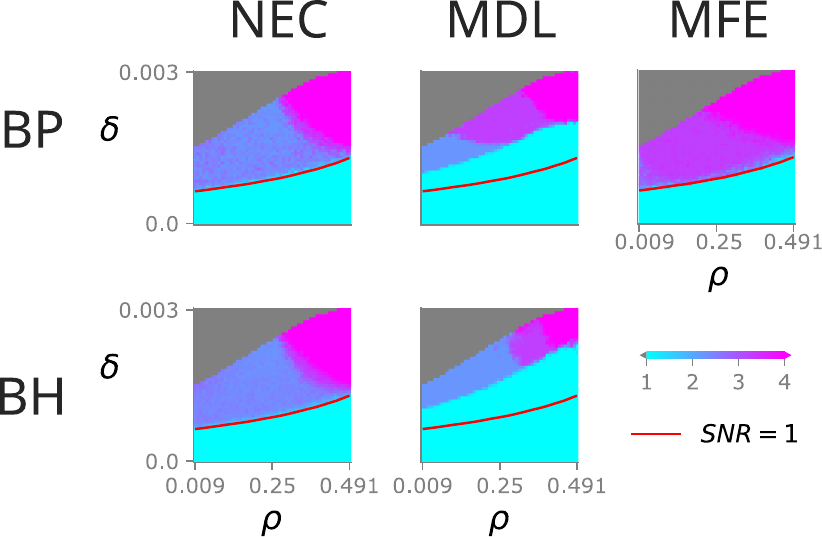}\\
	\caption{\small\it The experimental results reveal distinct phase transitions above the detectability threshold $\rm SNR=1$, marked by the red line, across different combinations of community detection and order selection methods.
    All experiments adopt the same parameter setting under the consistent $\pout$ scenario: $n=6000$, $d=5$, $q_s=2$, $q_b=2$.
    Each subplot shows the average number of detected communities over multiple repetitions.
    Columns correspond to order selection methods: NEC, MDL, and MFE; rows indicate community detection methods: BP and BH.
    \label{fig3.2}}
\end{figure}
Figure~\ref{fig3.2} shows that different combinations of methods exhibit distinct phase transitions above the detectability threshold $\rm SNR=1$, with the gray area indicating invalid $(\rho, \delta)$ because $\pin, \pout\notin[0, 1]$. 
Key observations:
\begin{itemize}
    \item NEC (first column): Two distinct phase transitions above $\rm SNR=1$, progressing from $q=2$ to $q=3$ to $q=4$.
    \item MDL (second column): Communities are not detected all the way down to the detectability threshold due to the resolution limit of SBM using MDL without a hierarchical prior~\cite{peixoto2013parsimonious}; all phases remain observable but shifted.
    \item MFE (last column): Clear transition from $q=3$ to $q=4$, but the $q=2$ to $q=3$ transition is less sharp.
\end{itemize}

Order selection strategy significantly affects the observed phase boundaries (Fig.~\ref{fig3.2}).
The three-phase structure persists with BH-based spectral clustering, though transition thresholds vary across NEC, MFE, and MDL methods.
BP demonstrates comparable or superior performance to BH. 
However, unlike the BH results, BP with MFE order selection shows only a clear $q=2$ to $q=3$ transition (Fig.~\ref{fig3.2}, right panel). 
One possible explanation is that BP leverages degree heterogeneity between minority and majority communities in ways that spectral methods cannot. 
Under the consistent $\pout$ setting, these average degrees differ:
\begin{align}
        d_s & = \frac{n\rho}{q_s}(\pin-\pout) +n\pout \\
        d_b & = \frac{n(1-\rho)}{q_b}(\pin -\pout) + n\pout \enspace.
\end{align}

Since $\delta>0$ and $\epsilon<0$, we have $d_s < d_b$. This degree heterogeneity may provide an advantage for BP~\cite{zhang2016community}.

\subsection{Consistent average degree}\label{pt_mc_cd}
\begin{figure}
    \centering
    \includegraphics[width=\linewidth]{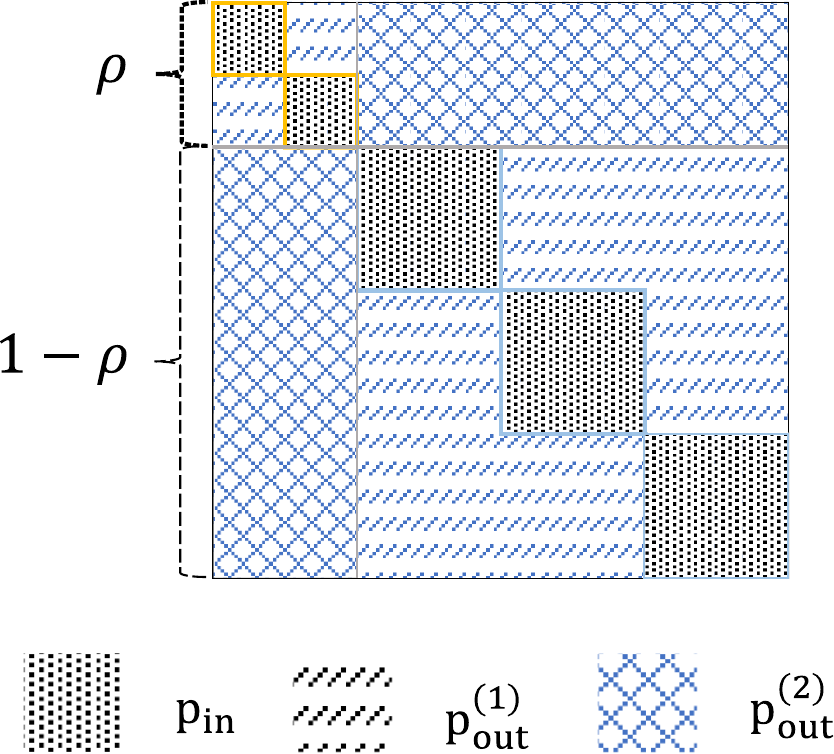}\\
	\caption{\small\it 
    An illustration of the expected adjacency matrix under consistent average degree setting.
    Here, two minority communities share a total proportion of $\rho$ nodes equally, while three majority communities share the remaining $1-\rho$ proportion equally. The patterns in different blocks represent the expected values of adjacency matrix entries, corresponding to the edge probabilities $\pin$, $\pout^{(1)}$, and $\pout^{(2)}$ in the affinity matrix of the SBM. \label{pout1_pout2}}
\end{figure}

To eliminate degree differences between communities of unequal size, we introduce a modified setting with two distinct inter-community edge probabilities, chosen to enforce equal average degrees across all communities. Figure~\ref{pout1_pout2} illustrates this configuration with the following parameters:
\begin{itemize}
    \item $\pin$ for the intra-community edge probability,
    \item $\pout^{(1)}$ for the inter-community edge probability within $\calS$ or within $\calB$,
    \item $\pout^{(2)}$ for the inter-community edge probability between $\calS$ and $\calB$.
\end{itemize}

For $\pout^{(1)}\neq \pout^{(2)}$, the average degree of minority communities $\calS$ and majority communities $\calB$ are given by:
\begin{align}\label{d_consistent}
    d_s&= \frac{n\rho}{q_s}(\pin-\pout^{(1)}) +n\rho\pout^{(1)}+n(1-\rho)\pout^{(2)}\\
    d_b&= \frac{n(1-\rho)}{q_b}(\pin -\pout^{(1)}) + n\rho\pout^{(2)}+n(1-\rho)\pout^{(1)} \enspace .
\end{align}

To ensure a consistent average degree across all communities, i.e., $d_b=d_s=d$, we choose the configuration that $q_s=2, q_b=3$, and $\rho\in (0, 0.4)$.
Under these constraints, the values of $\pin$, $\pout^{(1)}$, and $\pout^{(2)}$ are:
\begin{align}\label{pinpout1pout2}
    \pout^{(1)}&=\frac{d}{n}-\frac{1}{1-2\rho}\left(\frac{(1-\rho)^2}{q_b}-\frac{\rho^2}{q_s}\right)\delta \\
    \pout^{(2)}&=\frac{d}{n}+\frac{\rho(1-\rho)}{1-2\rho}\left(\frac{1}{q_b}-\frac{1}{q_s}\right)\delta \\
    \pin &= \frac{d}{n}+\left(1-\frac{1}{1-2\rho}\left(\frac{(1-\rho)^2}{q_b}-\frac{\rho^2}{q_s}\right)\right)\delta \enspace .
\end{align} 

The eigenvalues of $\bfQ$ under the consistent average degree setting, along with their derivation and ordering proof, are provided in Appendix~\ref{appendix_10_signal_spectrum_consistentd}. These eigenvalues define the identifiability threshold ($\lambda_3^2/\lambda_1=1$) and detectability threshold ($\lambda_4^2/\lambda_1=1$) for minority communities.

\begin{figure}
    \centering
    \includegraphics[width=\linewidth]{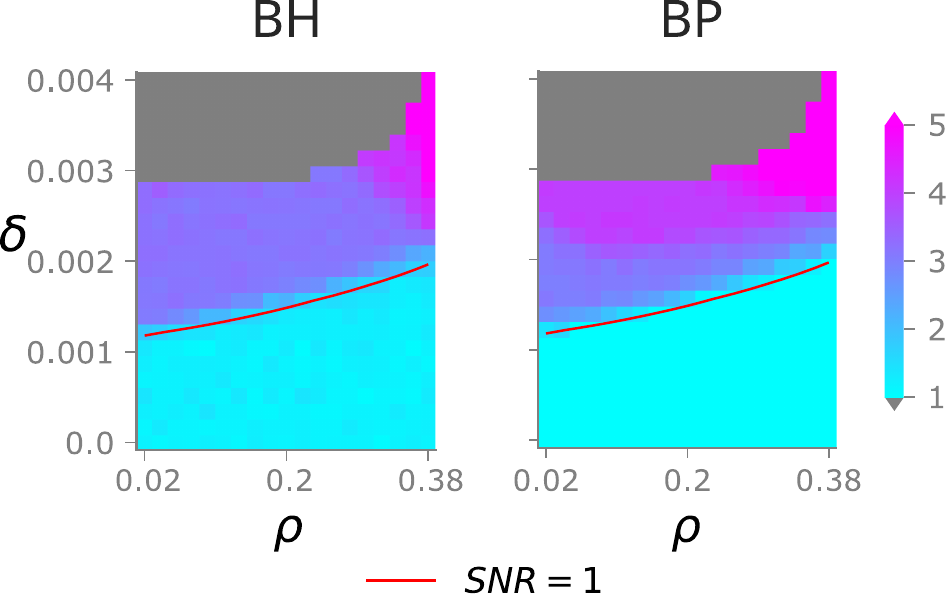}\\
	\caption{\small\it Experimental results under the consistent average degree scenario with parameters $n=6000$, $d=5$, $q_s=2$, $q_b=3$.
    Each subplot shows the average number of detected communities over multiple repetitions.
    The left panel presents the results obtained using the \textbf{BH} method, while the right panel shows those from \textbf{BP} with order selection by \textbf{MFE}.\label{fig3.4}}
\end{figure}

Figure~\ref{fig3.4} compares BH and BP under the consistent average degree setting. Unlike the constant $\pout$ scenario in Fig.~\ref{fig3.2}, both methods exhibit all three phases (detectable, distinguishable, resolvable) here. However, phase transition boundaries differ between methods, with BP's transitions occurring earlier, consistent with BH being a linearized approximation of BP.

\section{Conclusion}
Here we establish that minority communities in SBM networks exhibit additional phase transitions beyond the standard Kesten-Stigum threshold. Using eigenvalue analysis of the signal matrix, we derived explicit conditions for three distinct phases under Bethe Hessian spectral clustering. The detectable phase corresponds to recoverable overall community structure, where minority communities are merged into majority groups. The distinguishable phase marks when minorities separate from the majority as a collective. The resolvable phase indicates when each minority community is individually recoverable. Our experiments confirm these predictions. We also observe that belief propagation exhibits the same three-phase structure, performing at least as well as spectral clustering across all regimes.

A key takeaway is that both belief propagation and Bethe Hessian spectral clustering achieve the theoretical detectability threshold for community detection, yet they diverge when it comes to identifying communities at finer resolutions. This discrepancy reveals that reaching the information-theoretic limit does not guarantee equivalent performance across all aspects of community recovery. We view these findings as opening several promising directions for future research. The observed gap between the two methods raises questions about the trade-offs between computational efficiency and detection accuracy. Spectral methods offer computational advantages, while belief propagation provides superior fine-grained resolution. One promising direction is hybrid algorithms that combine spectral initialization with lightweight refinement, or modified spectral embeddings that better capture size heterogeneity; such methods could achieve both efficiency and accuracy. These developments would strengthen our ability to detect minority communities in large real-world networks where small but significant groups can carry critical information.

\section{Acknowledgments}
The authors would like to thank John Andrew Fitzgerald for helpful conversations. 
JL was supported by the China Scholarship Council under grant number 202106380033. LP was supported in part by the Dutch Research Council (NWO) Talent Programme ENW-Vidi 2021 under grant number VI.Vidi.213.163.

\bibliography{reference}

\appendix
\section{Numerical Considerations for Free Energy-Based Order Selection}\label{appendix_4_round_free_energy_for_elbow_point}
The order selection algorithm presented in Section~\ref{chap3_bplearnq} relies on minimizing the free energy $f_q$ to determine the number of communities. However, statistical fluctuations and numerical imprecision can cause the observed minimum to shift, potentially leading to an overestimation of $q$. This appendix discusses the selection of comparison precision used to mitigate these effects.

\begin{algorithm}
    \caption{Order Selection by minimizing free energy}
    \label{chap3_bplearnq}
    \SetAlgoLined
    \KwData{Network $\calG$;\\
    The maximum possible number of communities $q^{\rm max}$;\\
    Repeat times for each $q$: $t$;
    }
    \KwResult{Number of communities $q^*$}
    \For{$q\in\left\{1, 2, ...,  q^{\rm max}\right\}$}{
        Run $t$ times \textbf{EM}
        on network $\calG$ to find the minimum free energy $f_q^{\rm min}$;\\
        Select an appropriate comparison precision of $f_q^{\rm min}$;
        \If{$q>1$ and $f_q^{\rm min}\geq f_{q-1}^{\rm min}$}
        {
            \Return{$q^*=q-1$}
        }
    }
    \Return{$q^*=q^{max}$}
\end{algorithm}

\begin{figure}[H]
    \centering
    \includegraphics[width=\linewidth]{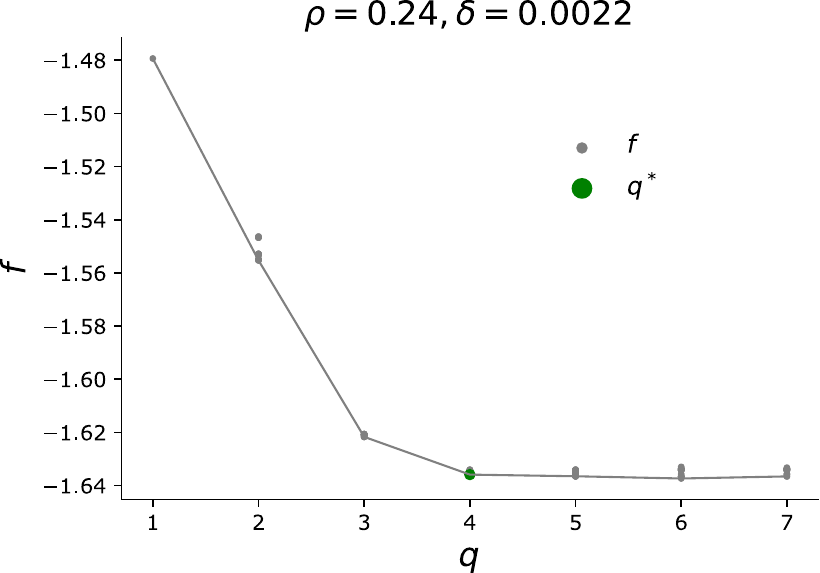}\\
	\caption{\small\it \textbf{BP} free energy as a function of the number of communities $q$, evaluated at $(\rho=0.24, \delta=0.0022)$. The parameters of the SBM with minority communities: $n=6000$, $d=5$, $q_s=2$, and $q_b=3$. \label{figa1}}
\end{figure}

As illustrated in Figure~\ref{figa1}, numerical noise can cause the apparent minimum to deviate from the true $q^*$.
For example, at the point $(\rho=0.24, \delta=0.0022)$ from the experiment illustrated in Figure~\ref{fig3.4}, we plot $f_q$ for $q\in[1, 7]$ in Figure~\ref{figa1}.
Visual inspection clearly suggests that $f_q^{\min}$ reaches a minimum at $q=4$ and then levels off, indicating $q^*=4$.
However, it is challenging to determine an appropriate precision for comparing $f_q^{\min}$ values. Minor differences in $f_q^{\min}$ beyond the $q^*=4$ can affect the selection of the final number of communities. 
In our experimental setup, after reviewing free energy plots across several $(\rho, \delta)$ points, we opt to retain three decimal places for $f_q^{\min}$ when the candidate $q < 3$, and two decimal places when $q \geq 3$. Here the threshold is placed on the candidate $q$ being tested in the loop, not on $q^*$, since the latter is unknown a priori.

\section{Condition for the constant-mode eigenvalue to become an eigenvalue of $\bfQ/n$} \label{appendix_5_lambda1_hat_bfA}
The minority submatrix $\frac{\rho}{q_s}\bfOmega_{q_s}$ has two distinct eigenvalues: the community-contrast eigenvalue $\lambda_4$, which extends to a full-matrix eigenvalue of $\bfQ/n$ with multiplicity $q_s-1$, and the constant-mode eigenvalue $\lambda_1^{\hat{\bfA}}$, corresponding to the uniform all-ones vector within the minority block. Here we derive the conditions under which $\lambda_1^{\hat{\bfA}}$ also extends to an eigenvalue of $\bfQ/n$, and show that in all such cases it coincides with an eigenvalue already present in the spectrum.

Here, $\lambda_1^{\hat{\bfA}}=\frac{\rho}{q_s}(\pin-\pout+q_s\pout)$ denotes the constant-mode eigenvalue on the minority block (the larger eigenvalue of $\frac{\rho}{q_s} \bfOmega_{q_s}$). This eigenvalue extends to an eigenvalue of $\bfQ/n$ only under specific parameter conditions. To find these conditions, we examine whether the determinant of $\bfQ/n - \lambda_1^{\hat\bfA} \bfI$ equals zero.
\begin{equation}\label{PQ-lambdaA1I}
    \bfQ/n - \lambda_1^{\hat\bfA} \bfI =
    \begin{pmatrix}
        \hat\bfA & \hat\bfB \\
        \hat\bfC & \hat\bfD
    \end{pmatrix} \enspace ,
\end{equation}
where the blocks, evaluated at $\lambda = \lambda_1^{\hat\bfA}$, are
\begin{equation}\label{blocks_appendixB}
    \begin{aligned}
        \hat\bfA &= \bigl[\tfrac{\rho}{q_s}(\pin-\pout) - \lambda_1^{\hat\bfA}\bigr]\bfI_{q_s} + \tfrac{\rho}{q_s}\pout\bfJ_{q_s} \enspace ,\\
        \hat\bfB &= \tfrac{\rho}{q_s}\pout\bfJ_{q_s\times q_b} \enspace ,\\
        \hat\bfC &= \tfrac{1-\rho}{q_b}\pout\bfJ_{q_b\times q_s} \enspace ,\\
        \hat\bfD &= \tfrac{1-\rho}{q_b}\bfOmega_{q_b} - \lambda_1^{\hat\bfA}\bfI_{q_b} \enspace .
    \end{aligned}
\end{equation}
Because the determinant does not change with elementary column addition operations, we perform some elementary column addition operations on equation~\eqref{PQ-lambdaA1I}. First, we subtract the first column from every column except the first one. Then we get an arrow-shaped matrix~\eqref{arrowPQ-lambdaA1I} whose only non-zero elements are in the first row, the first column, and the diagonal.
\begin{widetext}
\begin{equation}\label{arrowPQ-lambdaA1I}
    \begin{aligned}
        &\left[
        \begin{array}{cccccc}
            (\frac{\rho}{q_s}-\rho)\pout&...&\rho\pout&\rho\pout&...&\rho\pout \\
            ...& \ddots & ...&...&...&...\\
            \frac{\rho}{q_s} \pout&...& -\rho\pout&0&...&0\\
            \frac{1-\rho}{q_b}\pout& ...& 0&\substack{(\frac{1-\rho}{q_b}-\frac{\rho}{q_s})\delta-\\\rho\pout}&...& 0\\
            ...& ...& ...&...&\ddots&...\\
            \frac{1-\rho}{q_b}\pout& ...& 0& 0&... & \ddots
        \end{array}
        \right]_{q\times q}\enspace .
    \end{aligned}
\end{equation}
\end{widetext}
Then we perform some elementary column addition operations to make all elements in the first column except the first element equal to 0. For simplicity, we do not show the final matrix. The determinant of the final matrix is the product of its diagonal elements, which is 
\begin{equation}
    \begin{aligned}
         \det(\bfQ/n & - \lambda_1^{\hat\bfA} \bfI)\\
         =&(-\rho\pout)^{q_s-1}\times\left[ \left(\frac{1-\rho}{q_b}-\frac{\rho}{q_s}\right)\delta-\rho\pout \right]^{q_b}\times\\
         &\left[ \left(\frac{\rho}{q_s}-\rho\right)\pout-\rho\pout\frac{\rho\pout / q_s}{-\rho\pout}(q_s-1)\right.\\
         &\left.-\rho\pout\frac{(1-\rho)\pout / q_b}{\left(\frac{1-\rho}{q_b}-\frac{\rho}{q_s}\right)\delta-\rho\pout}q_b \right]\\
         =&(-\rho\pout)^{q_s-1}\times\left[ \left(\frac{1-\rho}{q_b}-\frac{\rho}{q_s}\right)\delta-\rho\pout \right]^{q_b-1}\\
         &\times\rho(1-\rho)\pout^2\enspace .
    \end{aligned}
\end{equation}
Because $\rho\in (0, q_s/q)$, the only condition for $\det(\bfQ/n - \lambda_1^{\hat\bfA} \bfI)=0$ is $\pout=0$ or $(\frac{1-\rho}{q_b}-\frac{\rho}{q_s})\delta-\rho\pout=0$, which is equivalent to
\begin{equation}
    \lambda_1^{\hat\bfA}=\frac{1-\rho}{q_b}(\pin-\pout) \enspace .
\end{equation}

When $\pout=0$, $\bfQ/n$ possesses two distinct eigenvalues: $\lambda_4$ with multiplicity $q_s$ and $\lambda_2$ with multiplicity $q_b$.
In this case, the constant-mode eigenvalue coincides with $\lambda_4$.
Another case is when the constant-mode eigenvalue equals $\lambda_2$, which corresponds to the eigenvalue of $\bfQ/n$ with multiplicity $q_b-1$.
Since in both cases the constant-mode eigenvalue coincides with a pre-existing eigenvalue, we disregard it.

\section{Derivation of quadratic factor $Q(\lambda)$}\label{appendix_6}
Here we derive the explicit form of the quadratic factor $Q(\lambda)$, which appears in the factored characteristic polynomial of $\bfQ/n$ (equation~\eqref{Qdef}). The factor arises from the Schur complement of the minority block $\hat\bfA$ in $\bfQ/n - \lambda\bfI$; we compute it by inverting $\hat\bfA$ via the Sherman-Morrison formula and then evaluating the resulting determinant.

The quadratic factor $Q(\lambda)$ in equation~\eqref{Qdef} expands to the determinant expression in equation~\eqref{detD-CA-1B}, namely
\begin{equation}
    \det(\hat\bfD-\hat\bfC\hat\bfA^{-1}\hat\bfB) \enspace .
\end{equation}

We begin by computing the inverse of the matrix
\begin{equation}
    \hat\bfA=\left[\frac{\rho}{q_s}(\pin-\pout)-\lambda\right]\bfI_{q_s}+\frac{\rho}{q_s}\pout \bfJ_{q_s} \enspace .
\end{equation}

Using the Sherman-Morrison formula~\cite{petersen2008matrix}, which states that for an invertible matrix $\bfM$ and vector $\bm{u}$ and $\bm{v}$:
\begin{equation}\label{shermanmorrison}
    (\bfM+\bm{u}\bm{v}^T)^{-1}=\bfM^{-1}-\frac{\bfM^{-1}\bm{u}\bm{v}^T\bfM^{-1}}{1+\bm{v}^T\bfM^{-1}\bm{u}}\enspace ,
\end{equation}
we express $\hat\bfA$ in the form $a\bfI+b\mathbf{1}\cdot\mathbf{1}^T$, where
\begin{equation}
    \begin{aligned}
        a=\frac{\rho}{q_s}\delta-\lambda,
        b=\frac{\rho}{q_s}\pout
    \end{aligned}\enspace .
\end{equation}

Here, $\delta=\pin-\pout$ and $\mathbf{1}$ denotes the all-ones vector.
Applying the Sherman-Morrison formula~\eqref{shermanmorrison} on $\hat\bfA$ yields:
\begin{equation}
    \begin{aligned}
        \hat\bfA^{-1}
        &=\frac{1}{a}\bfI-\frac{\frac{1}{a}\bfI\cdot b\bfJ \cdot \frac{1}{a}\bfI}{1+\mathbf{1}^T\cdot \frac{1}{a}\bfI\cdot b\mathbf{1}}\\
        &=\frac{1}{a}\bfI-\frac{\frac{1}{a^2}b\bfJ}{1+\frac{b}{a}q_s}\\
        &=\frac{1}{a}\left(\bfI-\frac{b}{a+bq_s}\bfJ\right)\\
        &=\frac{1}{\frac{\rho}{q_s}\delta-\lambda}\left(\bfI_{q_s}-\frac{\frac{\rho}{q_s}\pout}{\frac{\rho}{q_s}\delta-\lambda+\rho\pout}\bfJ_{q_s}\right)
    \end{aligned} \enspace .
\end{equation}
From the block matrix structure defined in equation~\eqref{chap2_blockmatrix}, the submatrices are:
\begin{equation}
    \begin{aligned}
        \hat\bfB&=\frac{\rho}{q_s} \pout \bfJ_{q_s\times q_b}\\
        \hat\bfC&=\frac{1-\rho}{q_b}\pout \bfJ_{q_b\times q_s}\\
        \hat\bfD&=\frac{1-\rho}{q_b}\bfOmega_{q_b}-\lambda \bfI_{q_b}
    \end{aligned} \enspace .
\end{equation}

We now compute the product $\hat\bfC\hat\bfA^{-1}\hat\bfB$:
\begin{equation}
    \begin{aligned}
        &\hat\bfC\hat\bfA^{-1}\hat\bfB\\
        =&\frac{1-\rho}{q_b}\pout \bfJ_{q_b\times q_s}\hat{\bfA}^{-1}\frac{\rho}{q_s} \pout \bfJ_{q_s\times q_b}\\
        =&\frac{(1-\rho)\rho\pout^2}{q_bq_s}\frac{1}{\frac{\rho}{q_s}\delta-\lambda}\left(q_s-\frac{\frac{\rho}{q_s}\pout}{\frac{\rho}{q_s}\delta-\lambda+\rho\pout}q_s^2\right)\bfJ_{q_b\times q_b}\\
        =&\frac{(1-\rho)\rho\pout^2}{q_b}\frac{1}{\frac{\rho}{q_s}\delta-\lambda}\frac{\frac{\rho}{q_s}\delta-\lambda}{\frac{\rho}{q_s}\delta-\lambda+\rho\pout}\bfJ_{q_b\times q_b}\\
        =&\frac{(1-\rho)\rho\pout^2}{q_b}\frac{1}{\frac{\rho}{q_s}\delta-\lambda+\rho\pout}\bfJ_{q_b\times q_b}
    \end{aligned}\enspace ,
\end{equation}
Next, we evaluate the expression inside the determinant:
\begin{equation}
    \begin{aligned}
        &\hat\bfD-\hat\bfC\hat\bfA^{-1}\hat\bfB\\
        =&\frac{1-\rho}{q_b}(\delta\bfI_{q_b}+\pout \bfJ_{q_b})-\lambda \bfI_{q_b}-\\
        &\frac{(1-\rho)\rho\pout^2}{q_b}\frac{1}{\frac{\rho}{q_s}\delta-\lambda+\rho\pout}\bfJ_{q_b} \\ 
        =&\left(\frac{1-\rho}{q_b}\delta-\lambda\right)\bfI_{q_b}+\\
        &\left(\frac{1-\rho}{q_b}\pout-\frac{(1-\rho)\rho\pout^2}{q_b}\frac{1}{\frac{\rho}{q_s}\delta-\lambda+\rho\pout}\right)\bfJ_{q_b}\\
        =&\left(\frac{1-\rho}{q_b}\delta-\lambda\right)\bfI_{q_b}+\frac{(1-\rho)\pout}{q_b}\frac{\frac{\rho}{q_s}\delta-\lambda}{\frac{\rho}{q_s}\delta-\lambda+\rho\pout}\bfJ_{q_b}
    \end{aligned}\enspace.
\end{equation}
Since the determinant of a matrix equals the product of its eigenvalues, we compute the eigenvalues of $\hat\bfD-\hat\bfC\hat\bfA^{-1}\hat\bfB$. This yields:
\begin{equation}
    \begin{aligned}
        \det(\hat\bfD &-\hat\bfC\hat\bfA^{-1}\hat\bfB)\\
        =&\left(\frac{1-\rho}{q_b}\delta-\lambda\right)^{q_b-1}\\
        &\times\left(\frac{1-\rho}{q_b}\delta-\lambda+\frac{(1-\rho)\pout(\frac{\rho}{q_s}\delta-\lambda)}{\frac{\rho}{q_s}\delta-\lambda+\rho\pout}\right)\\
        =&\left(\frac{1-\rho}{q_b}\delta-\lambda\right)^{q_b-1}\times\frac{1}{\frac{\rho}{q_s}\delta+\rho\pout-\lambda}\times\\
        &\left[ \left(\frac{1-\rho}{q_b}\delta-\lambda\right)\left(\frac{\rho}{q_s}\delta+\rho\pout-\lambda\right)\right.\\
        &\left.\quad+ (1-\rho)\pout\left(\frac{\rho}{q_s}\delta-\lambda\right)\right] \\
        =& \left(\frac{1-\rho}{q_b}\delta-\lambda\right)^{\!q_b-1}
        \frac{Q(\lambda)}{\frac{\rho}{q_s}\delta+\rho\pout-\lambda}
    \end{aligned}\enspace .
\end{equation}

\section{Alternative method for deriving eigenvalues of $\bfQ/n$} \label{appendix_7_anotherderiving}
Here we derive the eigenvalues of $\bfQ/n$ using an alternative approach to that of Appendix~\ref{appendix_6}. Rather than factoring via the Schur complement, we apply elementary column operations directly to $\bfQ/n - \lambda\bfI$ to reduce it to upper triangular form, then read off the eigenvalues as the diagonal entries.

We solve the characteristic equation
\begin{equation}\label{characteristiceq_}
    \det(\bfQ/n-\lambda \bfI)=0 \enspace .
\end{equation}
First, we perform some elementary column addition operations to transform $\bfQ/n - \lambda \bfI$ to an arrow-shaped matrix
\begin{equation}\label{arrowPQ-lambdaI}
    \begin{aligned}
        &\left[
        \begin{array}{ccccccc}
            \frac{\rho}{q_s}\pin-\lambda&-\frac{\rho}{q_s}\delta+\lambda &...&...&...&...&... \\ 
            \frac{\rho}{q_s}\pout&\frac{\rho}{q_s}\delta-\lambda &0&...&0&...&... \\
            \vdots&0&\frac{\rho}{q_s}\delta-\lambda &\vdots&\vdots&\vdots&\vdots \\
            \vdots& \vdots & ... & \ddots&\vdots&\vdots&\vdots\\
            \frac{1-\rho}{q_b} \pout&0&...&...& \frac{1-\rho}{q_b}\delta-\lambda&\vdots&\vdots\\
            \vdots& \vdots& ...&...&...&\ddots&\vdots\\
            \vdots& \vdots& ...&...&...&...&\ddots\\
        \end{array}
        \right] .
    \end{aligned}
\end{equation}

Then we perform some elementary column addition operations to transform~\eqref{arrowPQ-lambdaI} to an upper triangular matrix. For simplicity, we do not show the final matrix. The determinant of the final matrix is the product of its diagonal elements:
\begin{equation}\label{detailcharacteristiceq}
    \begin{aligned}
         \det\biggl(\frac{\bfQ}{n}  &  - \lambda \bfI \biggr)\\
         =&\left[\frac{\rho}{q_s}\delta-\lambda\right]^{q_s-1}\cdot\left[ \frac{1-\rho}{q_b}\delta-\lambda \right]^{q_b}\cdot\\
         &\left[ \frac{\rho}{q_s}\pin-\lambda
           -\frac{-\frac{\rho}{q_s}\delta+\lambda}{\frac{\rho}{q_s}\delta-\lambda}\frac{\rho}{q_s}\pout(q_s-1)\right.\\
         &\left.\quad-\frac{-\frac{\rho}{q_s}\delta+\lambda}{\frac{1-\rho}{q_b}\delta-\lambda}\frac{1-\rho}{q_b}\pout q_b \right]\\
         =&\left[\frac{\rho}{q_s}\delta-\lambda\right]^{q_s-1}\cdot\left[ \frac{1-\rho}{q_b}\delta-\lambda \right]^{q_b}\cdot\\
         &\left[ \frac{\rho}{q_s}\delta+\rho\pout-\lambda+\frac{\frac{\rho}{q_s}\delta-\lambda}{\frac{1-\rho}{q_b}\delta-\lambda}(1-\rho)\pout \right]\\
         =&\left[\frac{\rho}{q_s}\delta-\lambda\right]^{q_s-1}\cdot\left[ \frac{1-\rho}{q_b}\delta-\lambda \right]^{q_b-1}\cdot\\
         &\left[ \left(\frac{\rho}{q_s}\delta+\rho\pout-\lambda\right)\left(\frac{1-\rho}{q_b}\delta-\lambda\right)\right.\\
         &\left.\quad+\left(\frac{\rho}{q_s}\delta-\lambda\right)(1-\rho)\pout \right]\\
         =&\left[\frac{\rho}{q_s}\delta-\lambda\right]^{q_s-1}\cdot\left[ \frac{1-\rho}{q_b}\delta-\lambda \right]^{q_b-1}\cdot\\
         &\left[ \lambda^2-\left(\left(\frac{\rho}{q_s}+\frac{1-\rho}{q_b}\right)\delta+\pout\right)\lambda+ \right.\\
         &\left.(\frac{\rho}{q_s}\delta+\rho\pout)\frac{1-\rho}{q_b}\delta+\frac{\rho}{q_s}\delta(1-\rho)\pout\right]
         \enspace .
    \end{aligned}
\end{equation}

The characteristic equation~\eqref{characteristiceq_} therefore has root $\lambda_4 = \frac{\rho}{q_s}\delta$ (equation~\eqref{lambda_4_definition}) with multiplicity $q_s-1$, and $\lambda_2 = \frac{1-\rho}{q_b}\delta$ with multiplicity $q_b-1$. The remaining two roots $\lambda_1$ and $\lambda_3$ are solutions to the quadratic in the last factor of equation~\eqref{detailcharacteristiceq}, whose coefficients match those of equation~\eqref{coefficients}. This confirms that both derivation methods yield the same eigenvalues.

\section{Generation of networks with minority communities}\label{appendix_8_network_generation}
\begin{figure}
    \centering
    \includegraphics[width=\linewidth]{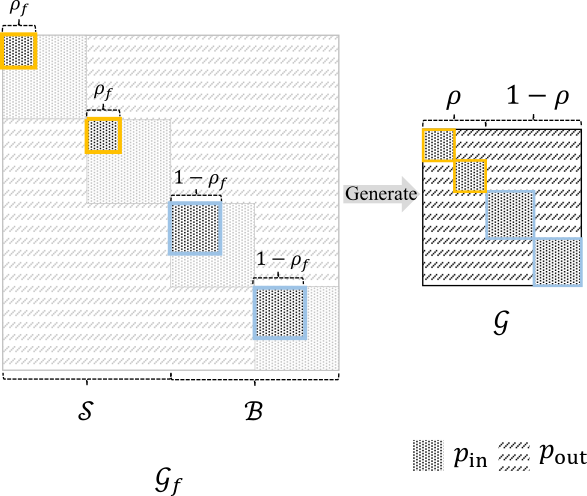}\\
	\caption{\small\it Constructing a network $\calG$ with minority communities by sampling from a larger symmetric SBM network $\calG_f$. \label{figa2}}
\end{figure}

Here we describe the method for generating synthetic networks under the consistent $\pout$ scenario. 
The network $\calG$ is constructed by sampling a subgraph from a larger background network $\calG_f$, which is generated from a symmetric SBM with $n_f$ nodes and $q = q_s + q_b$ communities.
In $\calG_f$, edges within communities occur with probability $\pin$, and edges between communities with probability $\pout$, as illustrated in Figure~\ref{figa2}.

After generating $\calG_f$, we form $\calG$ by selecting nodes as follows: from each of $q_s$ designated minority communities, we sample a fraction $\rho_f$ of nodes;
from each of the remaining $q_b$ communities, we sample a fraction $1-\rho_f$ of the nodes.
To ensure that the final network $\calG$ contains $n$ nodes and a proportion $\rho$ of nodes in minority communities, we set
\begin{equation}
    \begin{aligned}
        \rho_f &= \frac{q_b\rho}{q_b\rho+q_s(1-\rho)}\\
        n_f&= n \frac{(q_s+q_b)(q_b\rho+q_s(1-\rho))}{q_sq_b}
    \end{aligned}\enspace .
\end{equation}

Let $d$ denote the average degree of network $\calG$. We derive $d$ as
\begin{equation}\label{appendix_8_d}
    \begin{aligned}
        d=&\frac{1}{n}\left\{n(1-\rho)\left[ \frac{n(1-\rho)}{q_b}(\pin+(q_b-1)\pout)+n\rho\pout \right] + \right.\\
        &\left. n\rho\left[ n(1-\rho)\pout + \frac{n\rho}{q_s}(\pin+(q_s-1)\pout) \right] \right\}\\
        =&n\left[ \left(\frac{(1-\rho)^2}{q_b}+\frac{\rho^2}{q_s}\right)\pin+ \left( 1-\frac{(1-\rho)^2}{q_b}-\frac{\rho^2}{q_s} \right)\pout \right]
    \end{aligned}\enspace .
\end{equation}

Given that $\delta=\pin-\pout$, we can express $\pin, \pout$ in terms of $d$ and $\delta$:
\begin{equation}\label{appendix_8_pinpout}
    \begin{aligned}
        \pin&=\frac{d}{n}-\left(\frac{(1-\rho)^2}{q_b}+\frac{\rho^2}{q_s}\right)\delta+\delta \\
        \pout&=\frac{d}{n}-\left(\frac{(1-\rho)^2}{q_b}+\frac{\rho^2}{q_s}\right)\delta 
    \end{aligned}\enspace .
\end{equation}

To ensure $\pin, \pout\in[0, 1]$, $\delta$ must lie within the following feasible range:
\begin{equation}\label{appendix_8_deltarange}
    \begin{aligned}
        \frac{\frac{d}{n}-1}{\frac{(1-\rho)^2}{q_b}+\frac{\rho^2}{q_s}} \leq &\delta \leq \frac{\frac{d}{n}}{\frac{(1-\rho)^2}{q_b}+\frac{\rho^2}{q_s}} \\
        \frac{\frac{d}{n}}{\frac{(1-\rho)^2}{q_b}+\frac{\rho^2}{q_s}-1} \leq &\delta \leq \frac{\frac{d}{n}-1}{\frac{(1-\rho)^2}{q_b}+\frac{\rho^2}{q_s}-1} 
    \end{aligned}\enspace .
\end{equation}
In practice, we can generate the network $\calG$ if $\delta$ lies within the joint feasible interval defined above~\eqref{appendix_8_deltarange}, as the supplementary experiments shown in Appendix~\ref{appendix_9}.

\section{Eigenvalues of signal matrix under consistent average degree}\label{appendix_10_signal_spectrum_consistentd}
Here we derive the eigenvalues of the signal matrix $\bfQ$ in the consistent average degree scenario introduced in Section~\ref{pt_mc_cd}. In this scenario the two inter-community edge probabilities $\pout^{(1)}$ and $\pout^{(2)}$ are distinct, so the block structure of $\bfQ$ differs from the consistent $\pout$ case. We apply the block determinant formula and Sherman-Morrison inversion to factor the characteristic polynomial, then establish the ordering of the four eigenvalues. We also derive the eigenvalues for the case where all majority communities are merged into one.

In the consistent average degree scenario, where $\delta=\pin-\pout^{(1)}$, the signal matrix is:
\begin{equation}
\begin{aligned}
\bfQ=&n\left [
\begin{array}{c|c}
\frac{\rho}{q_s}\bfOmega^{(1)}_{q_s}& \frac{\rho}{q_s}\pout^{(2)} \bfJ_{q_s\times q_b}  \\
        \hline
        \frac{1-\rho}{q_b}\pout^{(2)} \bfJ_{q_b\times q_s} & \frac{1-\rho}{q_b}\bfOmega^{(1)}_{q_b}\\
\end{array}
\right]_{q\times q}\\
\end{aligned} \enspace ,
\end{equation}
where $\bfOmega^{(1)}_{q} = (\pin-\pout^{(1)})\bfI_{q}+\pout^{(1)}\bfJ_{q}$ is the community structure matrix using $\pout^{(1)}$.

To compute the eigenvalues of $\bfQ$, we need to find the roots of the characteristic polynomial $\frac{1}{n}\bfQ-\lambda \bfI$, which can be written in block-matrix form:
\begin{equation}
    \begin{pmatrix} \hat{\bfA} & \hat{\bfB}\\ \hat{\bfC} & \hat{\bfD} \end{pmatrix} \enspace ,
\end{equation}
with the blocks defined as:
\begin{equation}
\begin{aligned}
\hat{\bfA} =& \left(\frac{\rho}{q_s}\delta-\lambda \right)\bfI_{q_s}+\frac{\rho}{q_s}\pout^{(1)}\bfJ_{q_s}\\
\hat{\bfB} =& \frac{\rho}{q_s}\pout^{(2)} \bfJ_{q_s\times q_b}\\
\hat{\bfC} =& \frac{1-\rho}{q_b}\pout^{(2)} \bfJ_{q_b\times q_s}\\
\hat{\bfD} =& \left(\frac{1-\rho}{q_b}\delta-\lambda \right)\bfI_{q_b}+\frac{1-\rho}{q_b}\pout^{(1)}\bfJ_{q_b}
\end{aligned} \enspace .
\end{equation}
Using the determinant formula for block matrices~\cite{silvester2000determinants}, we have:
\begin{equation}
    \det(\frac{1}{n}\bfQ-\lambda I)=\det(\hat{\bfA})\det(\hat{\bfD}-\hat{\bfC}\hat{\bfA}^{-1}\hat{\bfB}) \enspace .
\end{equation}
Applying the Sherman–Morrison formula~\cite{petersen2008matrix} as given in equation~\eqref{shermanmorrison}, we compute:
\begin{equation}
    \hat{\bfA}^{-1}=\frac{1}{\frac{\rho}{q_s}\delta-\lambda}\left(\bfI_{q_s}-\frac{\frac{\rho}{q_s}\pout^{(1)}}{\frac{\rho}{q_s}\delta-\lambda+\rho\pout^{(1)}}\bfJ_{q_s}\right) \enspace .
\end{equation}
Then, 
\begin{equation}
\begin{aligned}
\hat{\bfC}\hat{\bfA}^{-1}=&\frac{\frac{1-\rho}{q_b}\pout^{(2)}}{\frac{\rho}{q_s}\delta-\lambda}\left(\bfJ_{q_b\times q_s}-\right.\\
&\left.\frac{\rho\pout^{(1)}}{\frac{\rho}{q_s}\delta-\lambda+\rho\pout^{(1)}}\bfJ_{q_b\times q_s}\right)\\
=&\frac{\frac{1-\rho}{q_b}\pout^{(2)}}{\frac{\rho}{q_s}\delta-\lambda+\rho\pout^{(1)}}\bfJ_{q_b\times q_s}\\
\hat{\bfC}\hat{\bfA}^{-1}\hat{\bfB}=&\frac{\frac{\rho}{q_s}\pout^{(2)}\frac{1-\rho}{q_b}\pout^{(2)}q_s}{\frac{\rho}{q_s}\delta-\lambda+\rho\pout^{(1)}}\bfJ_{q_b}\\
=&\frac{\frac{\rho(1-\rho)}{q_b}\pout^{(2)}\pout^{(2)}}{\frac{\rho}{q_s}\delta-\lambda+\rho\pout^{(1)}}\bfJ_{q_b}\\
\hat{\bfD}-\hat{\bfC}\hat{\bfA}^{-1}\hat{\bfB}=&\left(\frac{1-\rho}{q_b}\delta-\lambda \right)\bfI_{q_b}+\\
&\left(\frac{1-\rho}{q_b}\pout^{(1)}-\frac{\frac{\rho(1-\rho)}{q_b}\pout^{(2)}\pout^{(2)}}{\frac{\rho}{q_s}\delta-\lambda+\rho\pout^{(1)}}\right)\bfJ_{q_b}
\end{aligned} \enspace .
\end{equation}
Since the determinant of a matrix equals the product of its eigenvalues, we obtain:
\begin{equation}
\begin{aligned}
\det(\hat{\bfA})=&\left(\frac{\rho}{q_s}\delta-\lambda \right)^{q_s-1}\left(\frac{\rho}{q_s}\delta-\lambda +\rho\pout^{(1)}\right)\\
\end{aligned} \enspace ,
\end{equation}
and
\begin{equation}\label{detD-CA-1B_consistentd}
    \begin{aligned}
        &\det(\hat{\bfD}-\hat{\bfC}\hat{\bfA}^{-1}\hat{\bfB})\\
        =&\left(\frac{1-\rho}{q_b}\delta-\lambda \right)^{q_b-1}\\
&\left(\frac{1-\rho}{q_b}\delta-\lambda +(1-\rho) \pout^{(1)}-\frac{\rho (1-\rho){\pout^{(2)}}^2}{\frac{\rho}{q_s}\delta-\lambda+\rho\pout^{(1)}}\right)
    \end{aligned} \enspace .
\end{equation}
Multiplying these gives the full characteristic polynomial:
\begin{equation}\label{det_consistend}
\begin{aligned}
&\det(\frac{1}{n}\bfQ-\lambda \bfI)\\
=&\left(\frac{1-\rho}{q_b}\delta-\lambda \right)^{q_b-1}\left(\frac{\rho}{q_s}\delta-\lambda \right)^{q_s-1}\cdot\\
&\left\{\left(\frac{1-\rho}{q_b}\delta-\lambda +(1-\rho)\pout^{(1)}\right)\left(\frac{\rho}{q_s}\delta-\lambda +\rho \pout^{(1)}\right)-\right.\\
&\left.\rho(1-\rho){\pout^{(2)}}^2  \right\}
\end{aligned}\enspace .
\end{equation}
The last term is a quadratic in $\lambda$: $a\lambda^2+b\lambda+c$, with coefficients:
\begin{equation}
\begin{aligned}
a =& 1\\
b =& -\left(\left(\frac{1-\rho}{q_b}+\frac{\rho}{q_s}\right)\delta+\pout^{(1)}\right)\\
c=&\left(\frac{1-\rho}{q_b}\delta+(1-\rho)\pout^{(1)}\right)\left(\frac{\rho}{q_s}\delta +\rho \pout^{(1)}\right)-\\
&\rho(1-\rho){\pout^{(2)}}^2
\end{aligned}\enspace .
\end{equation}
Based on~\eqref{pinpout1pout2}, the discriminant $\Delta=b^2-4ac$ is:
\begin{equation}
    \begin{aligned}
        \Delta=&b^2-4ac\\
        =&\left(\left(\frac{1-\rho}{q_b}\delta+(1-\rho)\pout^{(1)}\right)-\left(\frac{\rho}{q_s}\delta+\rho \pout^{(1)}\right)\right)^2+\\
        &4\rho(1-\rho){\pout^{(2)}}^2\\
        =&\left(\left(\frac{1-\rho}{q_b}-\frac{\rho}{q_s}\right)\delta+\right.\\
        &\left.(1-2\rho)\left(\frac{d}{n}-\frac{1}{1-2\rho}\left(\frac{(1-\rho)^2}{q_b}-\frac{\rho^2}{q_s}\right)\delta \right)\right)^2+\\
        &4\rho(1-\rho){\left(\frac{d}{n}+\frac{\rho(1-\rho)}{1-2\rho}\left(\frac{1}{q_b}-\frac{1}{q_s}\right)\delta\right)}^2\\
        =&\left((1-2\rho)\frac{d}{n}+\rho(1-\rho)\left(\frac{1}{q_b}-\frac{1}{q_s}\right)\delta\right)^2\left(1+\frac{4\rho(1-\rho)}{(1-2\rho)^2}\right)\\
        =&\left(\frac{d}{n}+\frac{\rho(1-\rho)}{1-2\rho}\left(\frac{1}{q_b}-\frac{1}{q_s}\right)\delta\right)^2
    \end{aligned}\enspace .
\end{equation}
Then the roots of this quadratic equation are:
\begin{equation}
\begin{aligned}
&\frac{1}{2}\left\{\left(\left(\frac{1-\rho}{q_b}+\frac{\rho}{q_s}\right)\delta+\pout^{(1)}\right)\pm\sqrt{\Delta} \right\}\\
=&\frac{1}{2}\left\{\left(\left(\frac{1-\rho}{q_b}+\frac{\rho}{q_s}\right)\delta+\pout^{(1)}\right)\pm\right.\\
&\left.\left(\frac{d}{n}+\frac{\rho(1-\rho)}{1-2\rho}\left(\frac{1}{q_b}-\frac{1}{q_s}\right)\delta\right)\right\}\\
=&\frac{1}{2}\left\{\left(\left(\frac{1-\rho}{q_b}+\frac{\rho}{q_s}\right)\delta+\frac{d}{n}-\frac{1}{1-2\rho}\left(\frac{(1-\rho)^2}{q_b}-\frac{\rho^2}{q_s}\right)\delta \right)\pm\right.\\
&\left.\left(\frac{d}{n}+\frac{\rho(1-\rho)}{1-2\rho}\left(\frac{1}{q_b}-\frac{1}{q_s}\right)\delta\right)\right\}\\
=&\frac{1}{2}\left\{\left(\frac{d}{n}+\frac{\rho(1-\rho)}{1-2\rho}\left(\frac{1}{q_s}-\frac{1}{q_b}\right)\delta\right)\pm\right.\\
&\left.\left(\frac{d}{n}+\frac{\rho(1-\rho)}{1-2\rho}\left(\frac{1}{q_b}-\frac{1}{q_s}\right)\delta\right)\right\}\\
=&\frac{d}{n} \text{ or }
\frac{\rho(1-\rho)}{1-2\rho}\left(\frac{1}{q_s}-\frac{1}{q_b}\right)\delta
\end{aligned}\enspace .
\end{equation}
Including the two eigenvalues from the first two factors of~\eqref{det_consistend}: $\frac{1-\rho}{q_b}\delta$ and $\frac{\rho}{q_s}\delta$, and multiplying all eigenvalues by $n$,
we obtain the eigenvalues of $\bfQ$:
\begin{equation}\label{appendix_eigvalue_consistentd}
\begin{aligned}
\lambda_1 =& d \\ 
\lambda_2 =& \frac{1-\rho}{q_b}n\delta \\
\lambda_3 =& \frac{\rho}{q_s}n\delta \\ 
\lambda_4 =& \frac{\rho(1-\rho)}{1-2\rho}\left(\frac{1}{q_s}-\frac{1}{q_b}\right)n\delta
\end{aligned} \enspace .
\end{equation}

For the case $\delta>0$ and $q_s<q_b$, and given the consistent average degree $d$ from~\eqref{d_consistent} and the minority condition $\epsilon<0$ from~\eqref{epsilon<0}, we have 
\begin{equation}
    \lambda_1>\lambda_2>\lambda_3\enspace .
\end{equation} 
Since $q_s<q_b$ and $\epsilon<0$, it follows that $\rho < \frac{q_s}{q}<\frac{1}{2}$, ensuring $\lambda_4>0$. The difference between $\lambda_4$ and $\lambda_3$ is:
\begin{equation}
    \begin{aligned}
        \lambda_4-\lambda_3 =& n\delta\left(\frac{\rho(1-\rho)}{1-2\rho}\frac{1}{q_s}-\frac{\rho(1-\rho)}{1-2\rho}\frac{1}{q_b}-\frac{\rho}{q_s}\right) \\
        =& n\delta \left(\frac{\rho^2}{1-2\rho}\frac{1}{q_s}-\frac{\rho(1-\rho)}{1-2\rho}\frac{1}{q_b} \right) \\
        =&n\delta\frac{1-\rho}{1-2\rho}\epsilon<0
    \end{aligned} \enspace .
\end{equation}
Therefore, the eigenvalues are ordered as:
\begin{equation}
    \lambda_1>\lambda_2>\lambda_3>\lambda_4 \enspace .
\end{equation}

\subsection*{A Heuristic Approach to Minority Community Detectability}
If we merge all majority communities into a single community while maintaining a consistent average degree, the intra-community edge probability of the merged majority community must be set as:
\begin{equation}
    \frac{\pin+(q_b-1)\pout^{(1)}}{q_b} \enspace .
\end{equation}
We continue to use the block matrix approach to derive the eigenvalues of $\bfQ$, modifying only the blocks $\hat\bfC$ and $\hat\bfD$ as follows:
\begin{equation}
    \begin{aligned}
        \hat{\bfC} =& (1-\rho)\pout^{(2)} \bfJ_{1\times q_s}\\
        \hat{\bfD} =& \left(\frac{1-\rho}{q_b}(\delta+q_b\pout^{(1)})-\lambda \right)
    \end{aligned} \enspace .
\end{equation}
Following a similar derivation, we find the first term in~\eqref{detD-CA-1B_consistentd} vanishes, while the remaining terms stay unchanged.
This yields the following set of eigenvalues:
\begin{equation}
    \begin{aligned}
    \lambda_1^{(m)} =& d \\ 
    \lambda_2^{(m)} =& \frac{\rho}{q_s}n\delta \\ 
    \lambda_3^{(m)} =& \frac{\rho(1-\rho)}{1-2\rho}\left(\frac{1}{q_s}-\frac{1}{q_b}\right)n\delta
    \end{aligned} \enspace .
\end{equation} 
We can observe that $\lambda_2^{(m)}$ in the merged community structure corresponds to $\lambda_3$ in~\eqref{appendix_eigvalue_consistentd} for the unmerged case.
This could be a hint that the detectability threshold of minority communities is related $\lambda_3$ of signal matrix $\bfQ$.

\section{Supplementary experiments on phase transition of minority communities}\label{appendix_9}
Here, we present additional experimental results to further validate the phase transitions described in Section~\ref{conjectured_pt} under the setting of a consistent inter-community edge probability $\pout$.
The experiments are conducted with parameters $n=6000$, $d=50$.
The interval of $\delta$ is carefully selected such that all $(\rho, \delta)$ pairs yield valid probabilities satisfying $\pin, \pout\in [0, 1]$.

We also examine disassortative cases where $\delta<0$, which further generalizes the detectability phase transitions identified in Section~\ref{conjectured_pt}. 
The results confirm that:
\begin{itemize}
    \item When $\rm SNR>1$ and $\frac{\lambda_3^2}{\lambda_1}<1$, the minority communities cannot be identified even as a single aggregated group;
    \item If $\frac{\lambda_3^2}{\lambda_1}>1$ and $\frac{\lambda_4^2}{\lambda_1}<1$, the minority communities become distinguishable as a distinct group but cannot be separated from one another;
    \item If $\frac{\lambda_4^2}{\lambda_1}>1$, the minority communities are fully resolvable and can be resolved individually.
\end{itemize}

\begin{figure*}
    \centering
    \begin{minipage}[b]{0.48\textwidth}
        \centering
        \includegraphics[width=\linewidth]{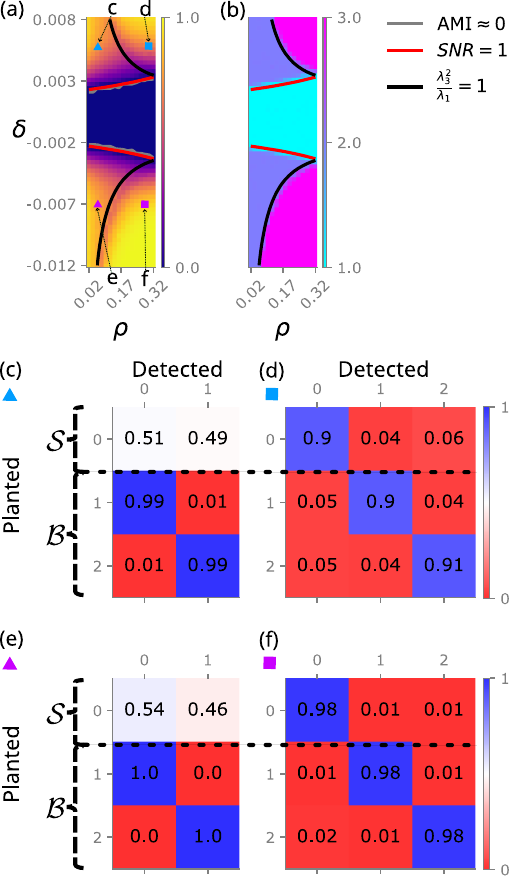}\\
    	\caption{\small\it Additional experimental results with $q_s=1$, $q_b=2$. For each $(\rho, \delta)$, the experiment was repeated 50 times.
        (a) The average $\rm AMI$. (b) The average number of detected communities. Both (a) and (b) are shown with curves $\mathrm{SNR}=1$ and $\frac{\lambda_3^2}{\lambda_1}=1$. (c--f) The confusion matrices for four representative $(\rho, \delta)$ points, (c):$(0.06, 0.006)$, (d):$(0.03, 0.006)$, (e):$(0.06, -0.007)$, and (f):$(0.03, -0.007)$. These matrices illustrate the phase transition behavior of minority communities across both assortative and disassortative cases.}
    \end{minipage}
    \hfill
    \begin{minipage}[b]{0.48\textwidth}
        \centering
        \includegraphics[width=\linewidth]{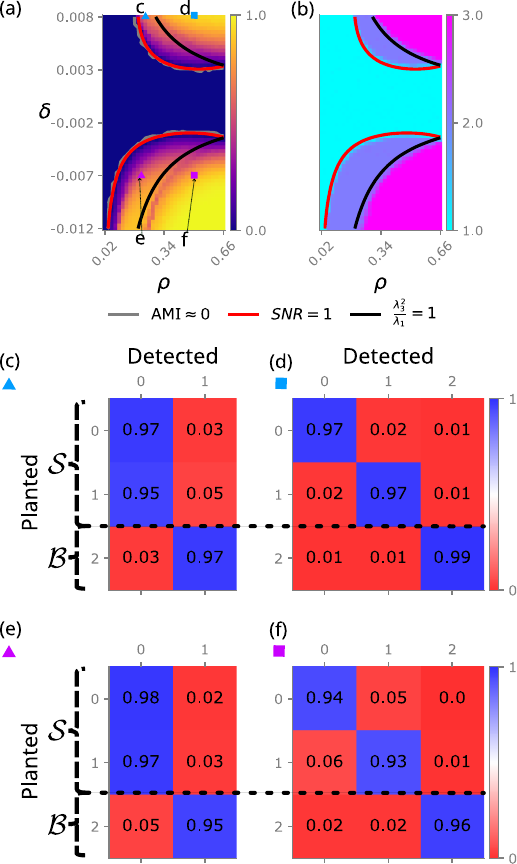}\\
    	\caption{\small\it Additional experimental results with $q_s=2$, $q_b=1$. For each $(\rho, \delta)$, the experiment was repeated 50 times.
        (a) The average $\rm AMI$. (b) The average number of detected communities. Both (a) and (b) are shown with curves $\mathrm{SNR}=1$ and $\frac{\lambda_3^2}{\lambda_1}=1$. (c--f) The confusion matrices for four representative $(\rho, \delta)$ points, (c):$(0.26, 0.008)$, (d):$(0.5, 0.008)$, (e):$(0.26, -0.007)$, (f):$(0.5, -0.007)$. }
    \end{minipage}
\end{figure*}

\begin{figure*}
    \centering
    \begin{minipage}[b]{0.48\textwidth}
        \includegraphics[width=\linewidth]{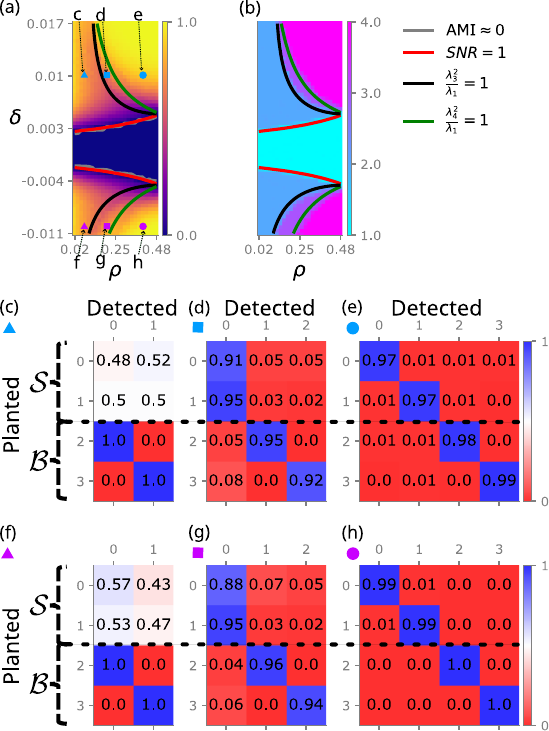}\\
    	\caption{\small\it Additional experimental results with $q_s=2$, $q_b=2$. For each $(\rho, \delta)$, the experiment was repeated 50 times. (a) The average $\rm AMI$. (b) The average number of detected communities. Both (a) and (b) are shown with curves $\mathrm{SNR}=1$, $\frac{\lambda_3^2}{\lambda_1}=1$, and $\frac{\lambda_4^2}{\lambda_1}=1$. (c--h) The confusion matrices for six representative $(\rho, \delta)$ points, (c):$(0.1, 0.01)$, (d):$(0.2, 0.01)$, (e):$(0.4, 0.01)$,  (f):$(0.1, -0.01)$, (g):$(0.2, -0.01)$, (h):$(0.4, -0.01)$. }
    \end{minipage}
    \hfill
    \begin{minipage}[b]{0.48\textwidth}
        \includegraphics[width=\linewidth]{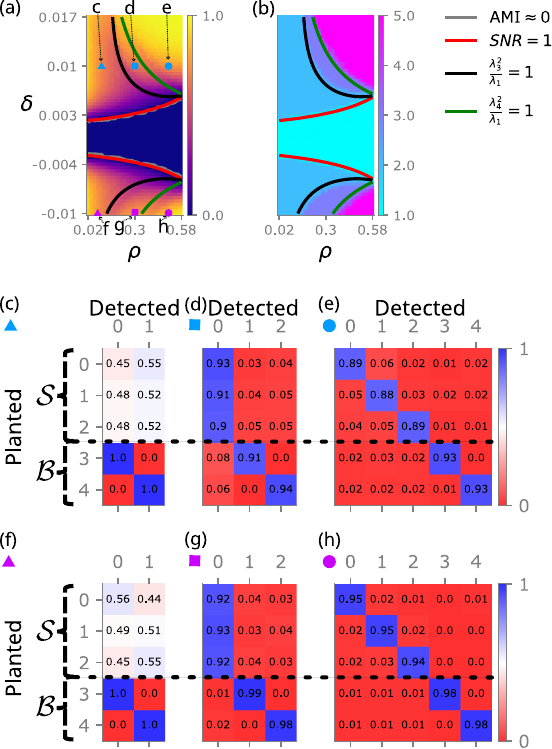}\\
    	\caption{\small\it Additional experimental results with $q_s=3$, $q_b=2$. For each $(\rho, \delta)$, the experiment was repeated 50 times. (a) The average $\rm AMI$. (b) The average number of detected communities. Both (a) and (b) are shown with curves $\mathrm{SNR}=1$, $\frac{\lambda_3^2}{\lambda_1}=1$, and $\frac{\lambda_4^2}{\lambda_1}=1$. (c--h) The confusion matrices for six representative $(\rho, \delta)$ points, (c):$(0.1, 0.01)$, (d):$(0.3, 0.01)$, (e):$(0.5, 0.01)$,  (f):$(0.1, -0.01)$, (g):$(0.3, -0.01)$, (h):$(0.5, -0.01)$.}
    \end{minipage}
\end{figure*}

\begin{figure}
    \centering
    \includegraphics[width=\linewidth]{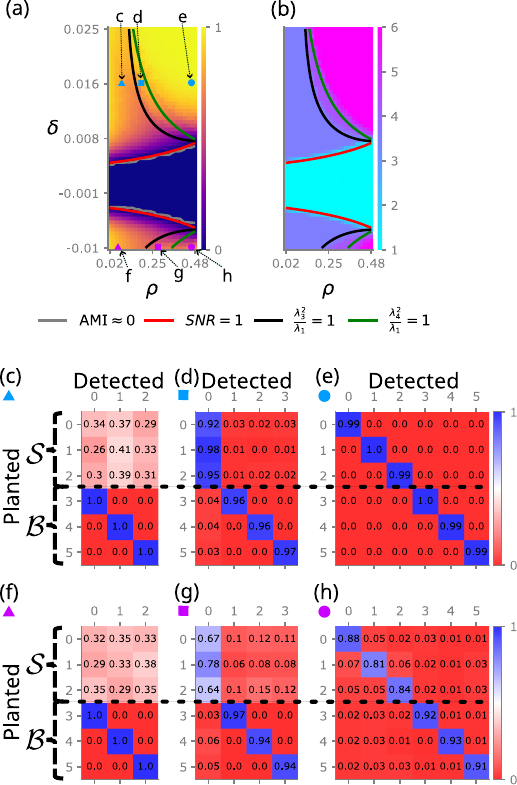}\\
	\caption{\small\it Additional experimental results with $q_s=3$, $q_b=3$. For each $(\rho, \delta)$, the experiment was repeated 50 times. (a) The average $\rm AMI$. (b) The average number of detected communities. Both (a) and (b) are shown with curves $\mathrm{SNR}=1$, $\frac{\lambda_3^2}{\lambda_1}=1$, and $\frac{\lambda_4^2}{\lambda_1}=1$. (c--h) The confusion matrices for six representative $(\rho, \delta)$ points, (c):$(0.1, 0.016)$, (d):$(0.18, 0.016)$, (e):$(0.46, 0.016)$,  (f):$(0.1, -0.01)$, (g):$(0.28, -0.01)$, (h):$(0.46, -0.01)$.}
\end{figure}

\end{document}